\documentstyle[12pt,fleqn]{article}
\renewcommand{\baselinestretch}{1.2}

\newcommand{\bea}{\begin{eqnarray}}
\newcommand{\beq}{\begin{equation}}
\newcommand{\eea}{\end{eqnarray}}
\newcommand{\eeq}{\end{equation}}
\newcommand{\nnu}{\nonumber}
\newcommand{\di}{\mbox{d}}
\newcommand{\dif}{\partial}

\newcommand{\spav}[1]{\parbox{1mm}{\vspace*{#1}}}

\begin{document}

\begin{titlepage}
\begin{flushright}
SISSA/ISAS 52--92--EP \\
April 1992
\end{flushright}
\spav{1cm}\\
\begin{center}
{\LARGE \bf \sc The Gravitational Field\\}
{\LARGE \bf \sc of String Matter}
\spav{1.5cm}\\
{\large  Marco Fabbrichesi, Roberto Iengo and Kaj Roland}
\spav{2cm}\\
{\em International School for Advanced Studies
(SISSA/ISAS)}\\
and \\
{\em INFN, Sezione di Trieste}
\spav{.5cm}\\
{\em via Beirut 2-4, I-34014 Trieste, Italy.} \\
\spav{2cm}\\
{\sc Abstract}
\end{center}
We study the scattering of a massless
and neutral  test particle in the
gravitational field of a body (the string star) made of a large number of
scalar  states of the
superstring. We consider two  cases, the one in which
these states are neutral string excitations
 massive already in ten dimensions  and the one in which
their masses (and charges)
originate in the process of compactification on tori.
 A perturbative calculation based on superstring amplitudes
gives us the deflection angle up to the second order in Newton's
constant. A comparison with field theory explicitly shows which among the
various massless fields of the superstring give a contribution to the
scattering process. In both cases, the deflection angle is smaller
than the one computed in general relativity. The perturbative series can
be resummed by finding the exact solution to the
 classical equations of
motion of the corresponding low-energy action. The
space-time metric of our two examples of
string stars  has no horizon.

\vfill
\end{titlepage}

\newpage
\setcounter{footnote}{0}
\setcounter{page}{1}

\section{Introduction}

\hspace*{.7cm}
 A {\em string star} is for us  a star which is made of
string matter, that is, of string states
 instead of ordinary matter like protons, neutrons and
electrons. We believe this to be
 a useful concept,
albeit only a hypothetical one,
in studying gravitation in string theory.

To be sure, in the framework of superstring theory, ordinary matter should
 be made out of the massless string states by compactification
and symmetry breaking through some
still unknown scenario. In this paper, however, we
leave these problems aside because
our purpose is to describe
 string matter in four space-time dimensions after
 compactification but before symmetry breaking
 occurs.

We
 consider two examples of string stars.

 The first example
  is the one in which string matter is  made
 of string excitations that are already massive in ten dimensions
 and are taken to be
  neutral with respect to  any gauge field, including those
 arising from the
compactification.
Such states
 can be thought of as a peculiar form of matter, not directly related to
 the one observable at low energies, but dominant
at very high energy densities and, in particular,
  near gravitational singularities~\cite{lavori}.
 All the results
 obtained in this case
   are completely independent of the compactification scheme.

 The second example we consider is the one
in which string matter is made of massive and charged string states that
were
 originally
 massless and neutral but
 acquired a mass
  and a charge by having a non-zero momentum in
 one of the compact directions.  We take all compact directions to be tori,
 thus leaving
  out those more sophisticated compactification
scenarios that are perhaps closer  to the experimental evidence
but for which the computation of string amplitudes becomes harder.

In order for the string star to be
an astronomical object it must be  made out of a very large
number of these massive  states. If we assume the string coupling constant to
be
of the order of $1/10$, the lowest excited state of the superstring has
a mass of the order of one hundredth of the Planck mass and
about $10^{40}$ of them  are
necessary to have a star with a mass comparable to our sun ($M_{\odot} \simeq 2
\times 10^{30}$ kg).

Much of our understanding of general relativity  relies
 on those examples of which we know an exact solution.
 Among these, the Schwarzschild solution~\cite{MTW}
 for the spherically symmetric field configuration in vacuum plays a
 preeminent role because it is a realistic
 description of the gravitational field of a star like our sun.

In string theory we  have a formal equivalence~\cite{GSW}
 between a piece of the low-energy
action of the superstring on the one hand and the Einstein-Hilbert action
of general relativity on the
other but an otherwise scarcity of explicit examples on which to train our
physical insight.

Recently,  interest in  exact solutions of string-inspired actions has
been revived~\cite{papers,harvey}, and similar questions
 have been  addressed, also
in comparison with results found in two-dimensional gravity.

Our approach is however  different:
we  start from a perturbative computation based on the
standard expression for the superstring
scattering amplitude in four space-time dimensions.
 We show
that the  amplitude for the scattering of a test
particle by the string star  is dominated by  the
expansion into tree diagrams, the legs of which are massless field
propagators. Since we can think of the test particle as
probing the metric, such an expansion is directly related to
the perturbative solution of the low-energy equations for the
 space-time metric and it can therefore
 be interpreted as a systematic expansion of the
metric in powers of the ratio of the gravitational radius of the star
over the impact parameter of the test particle
(a computation of that kind
was performed for the case of the Schwarzschild
metric  in ref.~\cite{Duff}). The
perturbative series can be resummed by finding an exact solution to
the classical equations of motions derived from the low-energy
effective action.

 Such a computation, based on classical equations, is
 consistent as long as, like in our case,
  the gravitational radius of the string star is much
 larger than the Planck length.

Section 2 of the paper contains a perturbative computation in which the
 scattering of a massless test particle  by the string star is computed
to the second order in Newton's constant. We develop a
systematic technique to handle  this kind of processes via  tree level on-shell
superstring amplitudes. In section 3, we compare
these results  with the corresponding ones in field theory. It is
shown that the amplitude can be interpreted in terms of field theory
Feynman diagrams where  the interaction is carried by massless fields. We can
thus identify which among these massless fields take part in the process.

In section
4 we consider directly the classical equations of motion of
the relevant massless fields,
as derived from the low-energy effective action.
We solve  these equations  exactly; at the lowest and next-to-the-lowest order
 the solutions reproduce the results obtained
by means of the superstring amplitude.
The  space-time metric is characterized by
 the absence of any
horizon  around the string star. This holds
in both cases, the
one in which  we
 consider neutral massive string excitations and the one in which
 we take string matter to be charged
 through compactification.  Otherwise the respective
space-time metrics differ  very much.

The case of the neutral massive string excitations
(also studied in ref.~\cite{FI}) gives a
solution which deviates from the standard Schwarzschild solution in that the
string excitations, although neutral with respect to gauge fields, are a
source for the dilaton field with a coupling which is
fixed by string perturbation theory.
Even though the massless test particle  interacts only with the
graviton---and therefore
the deflection is
 at the leading order the same as in
Einstein theory---the
metric is strongly affected  by the presence of the dilaton field, and
the solution belongs to the class of solutions of the
Brans-Dicke theory~\cite{BD} for a special choice of
the  parameters. As far as we know, this is a result which was not
previously discussed in the context of string theory.

The physical  relevance of such a scenario stems from several
studies~\cite{lavori} which indicate that, at high energy density, the
most probable configuration in string theory is the one in which most of
the massive states are
excited. Thus, one can speculate that an electrically neutral
collapsing star of ordinary matter and of sufficiently large mass would
eventually evolve into a string star like the one discussed in this paper.

The case of a star made of
originally  massless string states leads to a solution of the
bosonic sector of $N=8$ supergravity
in $D=4$
space-time dimensions. It belongs to the class of solutions of gravity
coupled to a scalar and a Maxwell field arising in the
compactification. It corresponds to the extremal
case discussed in~\cite{papers}, for which  the charge of the star is
proportional to its mass. In fact, it represents
an example of  anti-gravity~\cite{anti-gravity}, that is,
a theory in which the static gravitational
 interaction vanishes because the repulsive
 exchange of spin-1 fields compensates the attraction due to the
 graviton and the
spin-0 fields.
This is not  manifest in our case because we take the
massless test particle to be neutral and therefore
 insensitive to the vector and scalar
field exchange. The signature of the anti-gravity
solution is however present in the deflection angle where  non-linear
effects start only at the third order in Newton's constant.

 Both these  solutions  have a singularity of the  scalar curvature
 which is not
 hidden by a horizon. However, a quantum mechanical treatment
  of the  radial motion of a
 test particle  in the presence of the star
 indicates that  the particle would not fall  into the singularity.
 This is similar
 to the usual problem of the hydrogen atom where the kinetic energy
 (arising from the indeterminacy principle)
 prevents
 the electron from falling into the nucleus. One can compare this behavior
with the
 black hole case, where instead particles do fall into the horizon. This is
 discussed in section 5.

 In section 6 we put forward a possible argument suggesting  why our solution
 with neutral string states might be of relevance even in the more realistic
case in which the
 dilaton field becomes massive.

\section{The Perturbative Computation}
\setcounter{equation}{0}

\hspace*{.7cm} In this section we compute the
 deflection angle in the
classical scattering of a massless test particle
in the gravitational field of a star made of scalar string states.
We consider two possibilities for these states: either they are massive
already in ten dimensions or they acquire a mass in the process of
compactification. As we will see, the resulting theory of gravity is
very different in the two cases.

\subsection{Eikonal Multiple Scattering}

\hspace*{.7cm} In quantum field theory,
the problem of  scattering in a given external
field is usually treated  by computing Green functions in the presence
 of an external source. In physical terms, an incoming particle
 interacts with the background field by a multiple exchange of virtual
 particles.

In the case of string theory, we have no truly convenient second quantized
formulation; string amplitudes are well defined only on the mass
shell. A possible
procedure~\cite{FI} to compute the scattering of a particle
  in a background field by means
of on-shell amplitudes is the following.

 Consider
 the scattering of a massless test particle
by a distant target (a star of strings or  ordinary matter) made of
a large collection of $N$
scalar  states of mass $M$.
In a frame in which all the massive particles are
initially at rest, the massless one
moves  the same way a photon would in the gravitational field of  the
star.

Since all exchanged momenta are very small compared to the Planck mass,
the string amplitudes are dominated by contributions coming from corners
of moduli space that correspond to all possible exchanges
 of massless fields. It is thus possible to analyze the scattering in terms of
Feynman-like
diagrams in complete analogy with a field theory computation.
This point is further discussed in section
3.

To each order in Newton's constant
$G_N$, the leading contribution to the amplitude comes
from tree diagrams in which
the {\em trunk} of each massless field
 tree  splits into
{\em branches} attached to the
different scalar string states which make up the star, see figs. 1 and 2.
The tree diagrams with the external
states removed can be both connected (figs.1a and 1b) and
disconnected  (fig.1c).
\begin{figure}
\vspace{6cm}
\caption{Feynman-like diagrams relevant to the computation
of the deflection angle.}
\end{figure}
The dominance of tree over loop diagrams
 is true for combinatorial reasons alone. The number of tree-level
Feynman diagrams corresponding to a single tree with $m$ branches is
\beq
{N \choose m}
 \sim \frac{N^m}{m!} \:\:\:\: \mbox{for} \:\:\: N \rightarrow \infty\, ,
 \label{factor}
 \eeq
 whereas, for example, a one-loop diagram with $m$ branches can be realized
 in only
 \beq
 N {N-2 \choose m-2}
 \sim \frac{N^{m-1}}{(m-2)!}
\eeq
ways. We conclude that, in the limit $N \rightarrow \infty$,
 the computation of the scattering of a massless particle by such a
 composite target
is genuinely classical.

Furthermore, if we assume that all exchanged energies are small compared
to $M$, we can ignore the back-reaction of the particle on
the
constituents of the string  star and our computation will be equivalent to the
background field
method in quantum field theory. This means that
 the massless particle is a test particle
in the classical field generated by a stationary star and it is
possible to characterize the scattering by a single impact parameter
\beq
{\bf b} = \frac{{\bf L}}{|{\bf p}|} \, ,
\eeq
where {\bf L} is the angular momentum and ${\bf p}$ the momentum
 of the incoming test
particle. Moreover, the initial and final states of the star, $|\, i \rangle$
and
$|\, f \rangle$, are approximately equal. They can be expanded on free
particle states
\beq
|\, i \rangle = |\, f \rangle =  \int  \prod_{j=1}^{N} \left(
 \frac{\di ^3 {\bf p}_j}
{(2 \pi )^3 2E_j} \right)
\Psi (p_1, \ldots ,p_N ) |\, p_1, \ldots ,p_N \rangle
\, , \label{star-states}
\eeq
where the wave function $\Psi(p_1, \ldots ,p_N )$
describes an object localized in space
inside the surface of the string star and
with momenta distributed around the value
\beq
p_1 = \ldots = p_N = (M, {\bf 0})
\eeq
with a spread $|\Delta {\bf p} | \leq \Lambda$. The cutoff parameter
$\Lambda$ gives an upper bound on the allowed energy transfer in a single
exchange of a massless field and keeps the
scattering within the quasi-elastic regime.
In (\ref{star-states}), $E_j \simeq M$ denotes the energy of the $j$'th
particle.

It is convenient to
define the matrix element $A$ by factorizing in the $S$-matrix
 the normalization of the
external states of the test particle and the target:
\beq
\left( S - 1 \right) _{fi} = 2E\: \epsilon_{in} \cdot \epsilon_{out}
\: \sqrt{\langle i | i \rangle\langle f | f \rangle}
\: (2\pi i)  \:\delta (E_i + E_f )  A_{fi}  \, . \label{S}
\eeq

In Eq.~(\ref{S}), we have considered the case in which the massless
test particle in
the scattering process is a graviton; the $\epsilon$'s  are thus
the polarizations. This choice helps in being more specific; however,
our results apply also
to the case in which the graviton is replaced by any other massless state.
$E$ denotes the energy of the test particle.

By definition,
the $A$-matrix element is a transition amplitude for
free particle states averaged over the wave-function of the target:
\bea
A_{fi}  & \equiv   &
\int \frac{\di ^3 {\bf p}_1}
{(2 \pi )^3 2E_1} \int \frac{\di ^3 {\bf p}'_1}
{(2 \pi )^3 2E'_1} \cdots
\int \frac{\di ^3 {\bf p}_N}
{(2 \pi )^3 2E_N}
\int \frac{\di ^3 {\bf p}'_N}
{(2 \pi )^3 2E'_N} \nnu \\
& & \times
\Psi^* (p'_1, \ldots ,p'_N ) \Psi (p_1, \ldots ,p_N ) \nnu \\
& & \times (2\pi )^3
  \delta ^{(3)} ({\bf p}_{in} + {\bf p}_1 + \cdots
 + {\bf p}_N +{\bf p}_{out} +{\bf p}'_1 + \cdots + {\bf p}'_N) \nnu \\
& & \times
\frac{\langle p_{out}; p'_1,
 \ldots ,p'_N |\: T\:| p_{in}; p_1, \ldots ,p_N \rangle}
 {2E\: \epsilon_{in} \cdot \epsilon_{out}
  \sqrt{\langle i | i \rangle\langle f | f \rangle}} \label{T1}\, .
\eea

The $T$-matrix appearing in (\ref{T1}) is the usual free-particle
transition amplitude defined by
\bea
\langle p_{out}; p'_1,
 \ldots ,p'_N | &(S-1)& | p_{in}; p_1, \ldots ,p_N \rangle \\
 =
 & & i(2\pi)^4 \delta^{(4)} (\sum p_i )
 \langle p_{out}; p'_1,
 \ldots ,p'_N |\: T\:| p_{in}; p_1, \ldots ,p_N \rangle \nnu
 \eea
which can be computed   by means of
 Feynman diagrams in field theory and from the string path integral
in string
 theory.

Let us define
\beq
 q = p_{in} + p_{out}
\eeq
and for each exchange introduce the variables (see fig.2)
\bea
 q_i & = & - p_i - p'_i \\
 \Delta_i & = & \frac{1}{2} \left(  p_i -  p'_i\right)\, .
\eea
\begin{figure}
\vspace{6cm}
\caption{Parametrization of the tree diagram momenta.}
\end{figure}
By assumption it is possible to
 neglect the dependence of the  wave function of the target on
 the ${\bf
q}_i$'s; we can therefore factorize out in (\ref{T1}) the normalization factor
\beq
\sqrt{\langle i | i \rangle\langle f | f \rangle} =
\int \frac{\di ^3 {\bf \Delta}_1}
{(2 \pi )^3 2M} \cdots
\int \frac{\di ^3 {\bf \Delta}_N}
{(2 \pi )^3 2M} | \Psi ( \Delta_1, \ldots , \Delta_N ) |^2
\eeq
and obtain that
 \bea
 A_{fi} & = &
\int \frac{\di ^3 {\bf q}_1}
{(2 \pi )^3 } \cdots
\int \frac{\di ^3 {\bf q}_N}
{(2 \pi )^3 } \left( \frac{1}{2M} \right) ^N
(2\pi)^3 \delta^{(3)} ({\bf q}_1 + \cdots + {\bf q}_N - {\bf q}) \nnu \\
& & \times
\frac{\langle p_{out}; p'_1, \ldots ,p'_N |\:
T \:| p_{in}; p_1, \ldots ,p_N \rangle}{2E\: \epsilon_{in} \cdot
\epsilon_{out}}
\, .\label{T2}
\eea
 $T$ in eq.~(\ref{T2})  includes contributions from all
massless  field
exchanges, connected as well as disconnected.

In general, a connected $m$-branched tree-diagram will have $N-m$
spectator particles that only contribute a normalization factor
\beq
(2M)^{N-m} \prod_{j=1}^{N-m} \left[
(2\pi)^3 \delta^3({\bf q}_j) \right] \, .
\eeq
This fact allows us to
write the contribution from such a diagram
to the $A$-matrix element in its final form as
\bea
A^{(m)}_{fi}  &= &
\frac{N^m}{m!} \left( \frac{1}{2M} \right) ^m
\int \frac{\di ^3 {\bf q}_1}
{(2 \pi )^3 } \cdots
\int \frac{\di ^3 {\bf q}_m}
{(2 \pi )^3 }(2\pi)^3 \delta^{(3)}
 ({\bf q}_1 + \cdots + {\bf q}_m - {\bf q}) \nnu \\
& & \times
\frac{\langle p_{out}; p'_1, \ldots ,p'_m |\:
T^m \:| p_{in}; p_1, \ldots ,p_m \rangle}{2E\: \epsilon_{in} \cdot
\epsilon_{out}}
\label{T3} \, ,
\eea
where now $T^m$ refers to the particular connected diagram  in which
only $m$ scalar  massive states partake in the interaction and we have taken
 into account
 the combinatorial factor~(\ref{factor}).

Diagrams involving interactions between the constituents of the string
star can be consistently ignored in the computation of the $A$-matrix
element.

Insofar as we will deal only with processes in which the exchanged
momentum is kept fixed
and very much smaller than the energy of the incoming particle, it is
useful to cast our computation in the framework of
 the eikonal approximation~\cite{eikonal} in which the
 $A$-matrix for the
multiple scattering of the massless test particle on the composite
target can be written in the exponential form
\beq
A_{fi} =
\int {\mbox d}^{2} {\bf b} e^{i {\bf q}_{\bot} \cdot {\bf b}}
 \left[ \frac{ e^{i\delta ({\bf b})} -1}{i}
\right] \, ,  \label{T}
\eeq
where ${\bf q}_{\bot}$ is the transverse part of exchanged momentum (see
appendix D for a discussion of the kinematics).

At the lowest order in
$G_N$, the only contribution to the scattering
comes from the process in which the test particle exchanges a
single graviton with only one of the $N$ massive scalars
(Fig.1a); this can be
computed from the four-point
Veneziano amplitude~\cite{GSW} in which two gravitons and two massive scalars
are
taken as external states:
\beq
\langle p_{out}; p'_1 |T^1| p_{in}; p_1, \rangle =
\epsilon_{in} \cdot \epsilon_{out} \
\frac{4\kappa^2 M^2 E^2}{{\bf q}_{\bot}^2}  \, , \label{A4}
\eeq
where $\kappa$ is the gravitational coupling constant.

 Eq.~(\ref{A4}) can be used to
define the relationship between $\kappa$ and $G_N$ by
imposing that the deflection angle of the massless particle be equal to the one
obtained in general relativity.

To calculate the deflection angle $\Delta
\vartheta$ we use a stationary phase in (\ref{T}),
that is
\beq
\Delta \vartheta = -\frac{1}{E} \frac{\partial}{\partial b}
\delta ^{m} (b)  \, .\label{deflection-angle}
\eeq
For $T^1$ given by (\ref{A4}) we have
\beq
 \delta ^{m=1} (b)   =
 \int \frac{\di ^2 {\bf q}_{\bot}}{(2\pi)^2} e^{-i{\bf q_{\bot}}
\cdot {\bf b}}  \frac{\kappa^2 NME}{{\bf q}_{\bot}^2}
=  -\frac{\kappa^2 NME}{2\pi} \log b\label{delta-1} \, .
\eeq

Eq.~(\ref{deflection-angle}) and (\ref{delta-1}) yield the result
\beq
\Delta \vartheta = \frac{4M_{\displaystyle \star}}{b}
\frac{\kappa^2}{8\pi}    \, , \label{d}
\eeq
where $M_{\displaystyle \star} = NM$ is the total rest mass of the star.

The deflection (\ref{d}) is equal to the one in general
relativity~\cite{MTW}
if we define
\beq
\kappa^2 = 8\pi G_N   \, ,\label{k-G}
\eeq
which is the convention we keep throughout this paper.

An important point arises here. Definition (\ref{k-G}) is based on a
process in which only
 the graviton among the string massless fields is exchanged, the lower
 spin states being suppressed. Had we defined
the string coupling by means of the static potential between two bodies
of masses $M_1$ and $M_2$, by taking
\beq
V \equiv G_N \frac{M_1 M_2}{r}  \, ,
\eeq
the dilaton (and,
for charged string states,
other massless fields as well)
 would
have contributed. For instance, neutral, massive and
scalar string excitations
attract each other  by means of two forces,
one arising from the graviton and  one from the dilaton exchange,
the net result being an overall factor two with respect to the usual
Newtonian attraction.
This can be readily seen
by computing the four-point Veneziano amplitude for the external excited
string states. Accordingly,
 the static potential definition would have given us
$\kappa^2 = 4\pi G_N $ instead of (\ref{k-G}).

\subsection{The String Amplitude to the Second Order in $G_N$}

 \hspace*{.7cm} By counting powers of Newton's constant, the first
 non-linear correction to the deflection angle should be extracted from
 the six-point amplitude. Let us consider first the case
 in which the string scalar states of the star
  are massive already in ten dimensions (that is, they are string
  excitations).
 We perform the calculation of the superstring amplitude
by means of the covariant loop calculus~\cite{CLC}. In appendix A we
show that the same results can also be obtained  by using the
four-graviton string amplitude as a building block.

The six-point
amplitude for the scattering of a graviton off two massive scalars can be built
from the general operator vertex \cite{CLC,us}
\bea
\lefteqn{V_{6}  = \frac{4\pi^3}{\alpha ' \kappa^2}
\left\{
\int \frac{1}{\mbox{d}{\cal V}_{ABC}} \prod_{k=1}^{6}
\left( \mbox{d}Z_{k}
\ {}_k \! \langle q=0,0_{a}| \right)
\right.} \label{vertex}
 \\
&&   \!\!\!\!\! \times
\left. \exp \left[ \frac{1}{2} \sum_{k\neq l} \left(
\sqrt{\frac{\alpha '}{2}}p_{k} + \psi_{k}D_{Z_{k}} \right) \left(
\sqrt{\frac{\alpha '}{2}}p_{l}
+ \psi_{l}D_{Z_{l}} \right) \log ( Z_{k}-Z_{l} ) \right] \right\}
\wedge
\Biggl\{ c.c.\Biggr\} \nnu \, .
\eea

(\ref{vertex}) is written in the Lorentz-covariant world-sheet supersymmetric
formulation.
Locally on the super Riemann surface we choose holomorphic
super coordinates
\begin{equation}
Z=(z,\theta) \label{eq:Z}
\end{equation}
in terms of which the closed string space-time super coordinate can be
expanded into a bosonic and a fermionic part and into a left- and a
right-moving chirality sector as follows
\begin{equation}
X(Z,\bar{Z}) = x(z) + \bar{x}(\bar{z}) + i\theta  \psi(z) +
i\bar{\theta} \bar{\psi}(\bar{z}) \label{eq:X} \, ;
\end{equation}
these are then decomposed by the oscillator representation into
\bea
x(z) & = & \frac{1}{2} q - i \frac{\alpha ' }{2} p \log z +
i\sqrt{\frac{\alpha ' }{2}}\sum_{n \neq 0}
\frac{\alpha_{n}}{n}\, z^{-n}
\nnu \\
\psi(z) & = & \sqrt{\frac{\alpha ' }{2}}
\sum_{n \in {\bf Z}+1/2} \psi_{n}\, z^{-n-1/2} \label{3.1} \, ,
\eea
and similarly for the left mover. This decomposition can be performed in
the neighborhood of each puncture, thus introducing a set of
oscillators for each external state, as in (\ref{vertex}).

The differences in super space are defined as $Z-Y = z-y-\theta _{z}
\theta _{y}$. The covariant super derivative is defined as $D_{Z} \equiv
\partial_{\theta} + \theta \partial_{z}$. Finally, the states
$\langle q=0,0_{a}|$ are the position zero
vacuum states in the appropriate ghost sector.

 The super-projective invariant volume~\cite{CLC} is
\beq
d{\cal V}_{ABC}=\frac{dZ_{A} dZ_{B} dZ_{C}}
{ [( Z_{A}- Z_{B}) ( Z_{B}- Z_{C})
( Z_{C}- Z_{A}) ]^{1/2}}
\frac{1}{d\Theta_{ABC}}\; , \label{volume}
\eeq
where
\beq
\Theta_{ABC}=
\frac{\theta_A(Z_B-Z_C) + \theta_B(Z_C-Z_A) + \theta_C(Z_A-Z_B)
+ \theta_A\theta_B\theta_C}{[(Z_{A}-Z_{B})
(Z_{B}-Z_{C})(Z_{C}-Z_{A})
]^{1/2}} \; .\label{3.3}
\eeq
 The overall constant
appearing in front of~(\ref{vertex}) is the normalization of the
string amplitude on the sphere~\cite{Weinberg}.

We can fix the super-projective invariance
by choosing the following parametrization for
 the super Koba-Nielsen
variables
\bea
Z_{in} = (0,\theta_1) \:\:\:\: Z_2' & = & (y, \theta_y) \:\:\:\: Z_{1}'
 = (x, \theta_x)  \nnu\\
Z_{out} = (z, \theta_2)\:\:\:\: Z_2 &= &(1, 0) \:\:\:\: Z_1 = (\infty, 0)
\, .
\eea

The six-point $T$-matrix element is obtained
by acting with the vertex~(\ref{vertex}) on the external states of two
gravitons
\beq
\frac{\kappa}{\pi} \,\epsilon^{i}_{\mu} \bar{\epsilon}^i_{\bar{\mu}}
\psi_{-1/2}^{\mu} | p_{i} ;
 0 \rangle _{L}
\otimes  \overline{\psi}_{-1/2}^{\bar{\mu}}
| p_{i} ; 0 \rangle _{R} \label{graviton}
\eeq
(with $p_{i}^{2} = p_{i} \cdot \epsilon_{i} = p_{i}\cdot\bar{\epsilon}_{i}
= 0$),
and of four  scalars of mass $\alpha ' M^2 = 4$
\beq
\frac{\kappa}{3! \pi} \,\Xi_{i}^{\mu\nu\rho} \psi_{-1/2}^{\mu}
\psi_{-1/2}^{\nu}
\psi_{-1/2}^{\rho} |p_{i} ; 0 \rangle _{L} \otimes
\overline{\Xi}_{i}^{\bar{\mu}\bar{\nu}\bar{\rho}}
\overline{\psi}_{-1/2}^{\bar{\mu}} \overline{\psi}_{-1/2}^{\bar{\nu}}
\overline{\psi}_{-1/2}^{\bar{\rho}} | p_{i} ; 0 \rangle _{R} \label{scalar}
\eeq
(with $ p_{i}^{2} = -M^{2}$,
$\Xi_{i}^{\mu\nu\rho} = \epsilon^{\mu\nu\rho\lambda}p_{i}^{\lambda}/
\sqrt{3!}\: M$,
such that $\Xi_{i}^{\mu\nu\rho}\Xi^{i}_{\mu\nu\rho}=1$ and similarly for the
barred quantities). The numerical factor in front of both states are
fixed by the factorization of the amplitude~\cite{Weinberg}. No
 momentum is
flowing in the compact directions.

By considering only the part of the complete amplitude that is
proportional to $\epsilon_{in} \cdot \epsilon_{out}$,
 after integrating out the Grassman variables and performing
an integration by parts in $z$ to obtain a uniform power of $\alpha '$, we
find the following contribution to the
$T$-matrix
\bea
T_6 & = &
\epsilon_{in} \cdot \epsilon_{out} \frac{\kappa^4(\alpha ')^3}{4 \pi^3}
\,
\int \mbox{d}^{2}x \mbox{d}^{2}y
\mbox{d}^{2}z |1-y|^{-6+\alpha ' p_{2}'\cdot p_{2}} |y|^{\alpha
'p_{in}\cdot p_{2}'}
\nnu \\
& & \times |1-x|^{\alpha 'p_2 \cdot p_1'}|x|^{\alpha 'p_{in} \cdot p_1'}
|1-z|^{\alpha 'p_{out} \cdot p_2}|z|^{-2 + \alpha 'p_{in} \cdot p_{out}}  \nnu
\\
& & \times
|z-x|^{\alpha 'p_{out} \cdot p_1'} |z-y|^{\alpha 'p_{out} \cdot p_2'}
 |x-y|^{\alpha
'p_2' \cdot p_1'} \nnu \\
& &  \times \left\{
\frac{p_{out} \cdot p_1' \: p_2' \cdot p_1'}{(y-x)(z-x)}
+ \frac{p_{in} \cdot p_2' \: p_{out} \cdot p_1'}{y(z-x)}
 - \frac{p_{out} \cdot p_2' \: p_{in} \cdot p_1'}
{x(z-y)} \right. \nnu \\
& &
+ \frac{p_2' \cdot p_{out} \: p_2' \cdot p_1'}{(y-x)(z-y)}
 \left. + \frac{p_{out} \cdot p_2 \: p_2' \cdot p_1' }{(z-1)(y-x)}
\right\} \wedge \Biggr\{ c.c. \Biggr\} \, . \label{6}
\eea

Because of the three powers of
$\alpha '$ in front of $T_6$,
in order to get a non-vanishing result in the field theory limit $\alpha
' \rightarrow 0$ we have to extract three poles in the momenta---each of
them bringing down one power of $(\alpha')^{-1}$.
Since there are three internal propagators in a six-point $\phi^3$ tree
diagram, this is what
singles out the corners of moduli space corresponding to $\phi^3$-like
diagrams; each pole
comes from a
{\em pinching} limit in which some of the Koba-Nielsen variables come
close to each other.

Ignoring all diagrams describing interactions among string star constituents
there are three pinching limits we are interested in, as
depicted in fig.3.
\begin{figure}
\vspace{6cm}
\caption{The three pinched diagrams contributing to the
 deflection angle to  $O (G_N^2)$.}
\end{figure}

 The first one
corresponds to taking $z \rightarrow 0$ and, successively, $x \rightarrow
\infty$, $y \rightarrow 1$ (fig.3a). In terms of sewing parameters~\cite{us} we
have
$x=1/A_2$, $y=1-A_1$ and $z=A$; these variables are useful in evaluating
the amplitude in the pinching limit, since they all go to zero
at the same rate.  Rewriting (\ref{6})
in terms of the sewing parameters
pertaining to the diagram in fig.3a we arrive at:
\bea
T_6 & = &  \epsilon_{in} \cdot \epsilon_{out}
\frac{\kappa^4(\alpha ')^3}{4 \pi^3}
\,
\int \frac{\mbox{d}^{2}A \mbox{d}^{2}A_1
\mbox{d}^{2}A_2}{|A|^2|A_1|^6|A_2|^2}
 |A_2|^{-\alpha ' (p_{1}'\cdot p_{in} +
 p_{out}\cdot p_{1}' + p_2' \cdot p_1' +p_2 \cdot p_{1}')}
\nnu \\
& & \times
|A_1|^{\alpha ' p_2 \cdot p_2'} |A|^{ \alpha 'p_{in} \cdot p_{out}}
|1-A_1|^{\alpha' p_{in}\cdot p_2'}
|1-A_2|^{\alpha 'p_2 \cdot p_1'}
|1-A|^{\alpha 'p_{out} \cdot p_2}  \\
& & \times
|1 - AA_2|^{\alpha 'p_{out} \cdot p_1'} |1 -A-A_1|^{\alpha 'p_{out} \cdot p_2'}
 |1- A_2+A_2A_1|^{\alpha
'p_2' \cdot p_1'} \nnu \\
& &  \times \left\{
\frac{A_2 \: p_{out} \cdot p_1' \: p_2' \cdot p_1'}
{(1-AA_2)(1- A_2 +A_1A_2 )}
- \frac{p_{in} \cdot p_2' \: p_{out} \cdot p_1'}{(1-A_1)(1-AA_2)}
 + \frac{p_{out} \cdot p_2' \: p_{in} \cdot p_1'}
{(1 -A -A_1)} \right. \nnu \\
& &
+ \frac{p_2' \cdot p_{out} \: p_2' \cdot p_1'}{(1- A_2+A_1A_2 )
(1-A-A_1)}
 \left. + \frac{p_{out} \cdot p_2 \: p_2' \cdot p_1' }{(1-A)(1-A_2+A_1A_2)}
\right\} \wedge \Biggr\{ c.c. \Biggr\} \, .\nnu \label{6'}
\eea
By extracting the poles, we obtain (see appendix C)
\bea
T_{6}^{a} &=& \epsilon_{in} \cdot \epsilon_{out} \, \:
16 \kappa^4   \label{6a} \\
& & \times \frac{\left[ p_{in} \cdot p_1' \: p_{out} \cdot p_2' -
p_{in} \cdot p_2'\:  p_{out} \cdot p_1' +
 p_{out} \cdot p_2' \: p_2' \cdot p_1' +
  p_{out} \cdot p_2 \: p_{2}' \cdot p_1' \right]^2}
  {\left(  p_1' + p_1  \right) ^2 \left( p_{in} + p_{out}
  \right) ^2 \left( p_2' + p_2  \right) ^2 }
   \, .\nnu
\eea

After substitution of the momenta
 \bea
 p_{in} = (E, {\bf p}) & ; & p_{out} = (-E' , {\bf q} - {\bf p}) \nnu \\
 p_1 = p_2 & = & (M, {\bf 0}) \nnu \\
 p_1' &=& (-\widetilde{E}_1 , -{\bf q}_1) \nnu \\
 p_2' & = & (-\widetilde{E}_2 , -{\bf q}_2) \, ,\label{momenta}
 \eea
 where
 \beq
 \widetilde{E}_i = \sqrt{M^2 + {\bf q}^2_{i}} \simeq M +
 \frac{{\bf q}^2_{i}}{2M}\, ,
 \eeq
 and
\beq
{\bf q} = {\bf q}_1 + {\bf q}_2 \, ,
\eeq
eq.~(\ref{6a}) yields
\beq
T_6^a  = -\epsilon_{in} \cdot \epsilon_{out} \:
\frac{16 \kappa^4 M^{4} }
{{\bf q}_{\bot}^{2}\, {\bf q}_{1}^{2}\, {\bf q}_{2}^{2}}
({\bf q}_{1} \cdot {\bf p}) ({\bf q}_{2} \cdot {\bf p}) \label{6a'} \, .
\eeq

We keep in the numerator of (\ref{6a'}), and subsequent
formulas,
 only terms at most quadratic
in the momenta ${\bf q}$, ${\bf q}_1$
and ${\bf q}_2$, as higher order terms would not
give a contribution to the deflection angle.

The other two pinching limits are the one in
which $x= 1/A_2 $, $z=1-A$ and $y = 1- AA_1$ (the ladder, fig.3b) and the one
in which $x=1/(AA_2)$, $z=1/A$ and $y=1-A_1$ (the cross-ladder,
fig.3c). Notice
that now the Koba-Nielsen variables go to their pinched corners at
different rates.
These diagrams contain the iteration of the single-graviton exchange but also a
contact term where
the propagator separating the two graviton emissions is
 canceled by momenta factors in the vertices. Hence, we obtain
\bea
T_6^{b} &=& \epsilon_{in} \cdot \epsilon_{out} \, \:
16 \kappa^4 \nnu \\
& & \times \frac{\left[ p_{out} \cdot p_2 \: p_2' \cdot p_1' +
p_{in} \cdot p_1'\: p_{out} \cdot p_2' +
 p_{out} \cdot p_2'\: p_2' \cdot p_1' \right] ^2 }
 {\left(  p_1' + p_1  \right) ^2 \left( p_{out} + p_2 + p_2'
  \right) ^2 \left( p_2' + p_2  \right) ^2}
 \label{6b1}
\eea
and
\bea
T_6^{c}& &= \epsilon_{in} \cdot \epsilon_{out} \, \:
16 \kappa^4 \nnu \\
& & \times
\frac{\left[ p_{in} \cdot p_2'\: p_{out} \cdot p_1' \right] ^2 }
 {\left(  p_1' + p_1  \right) ^2 \left( p_{out} + p_1 + p_1'
  \right) ^2 \left( p_2' + p_2  \right) ^2 }\label{6b2}\, ,
\eea
respectively.

Again, by inserting the momenta (\ref{momenta}) we obtain
\beq
T_6^b + T_6^c =  \epsilon_{in} \cdot \epsilon_{out} \, \left\{
\frac{8\kappa ^{4} E^4 M^4}{
{\bf q}_{1}^{2} \, {\bf q}_{2}^{2}} \left[ \frac{1}
{{\bf q}_{1} \cdot {\bf p}} + \frac{1}{{\bf q}_{2} \cdot {\bf p}} \right]
+
\frac{16 \kappa ^{4} E^2 M^4}{{\bf q}_{1}^{2} \, {\bf q}_{2}^{2}} \right\}
\label{6b} \, .
\eeq

Notice that we need not include in our calculation
contact terms proportional to
${\bf q}_1^2 + {\bf q}_2^2$; they vanish after integration over ${\bf
q}_1$ and ${\bf
q}_2$  (see appendix C).

The first term in the amplitude~(\ref{6b}) is dominated by
momenta ${\bf q}_{1}$ orthogonal to ${\bf p}$ and gives rise to a
contribution to the $A$-matrix~(\ref{T3}) that can be written as
\beq
\left( A_{fi}^{(1)} \right)^2  =
i\frac{\kappa^4 M^2_{\displaystyle \star}
 E^2}{2} \int \frac{\di ^{2}{\bf q}_{1}
  d^{2}{\bf q}_{2}}{( 2 \pi )^2}
\frac{1}{{\bf q}_{1}^{2} \, {\bf q}_{2}^{2}}\:
\delta^{(2)} ({\bf q}_{1} +{\bf q}_{2}
- {\bf q}) \, .
\label{iter}
\eeq
The amplitude (\ref{iter}) is a convolution in momentum space of
two 4-point tree-level amplitudes.
As it has been discussed in~\cite{ACV}, such
a factorization implies the exponentiation of the lowest order contribution
typical of the eikonal approximation~\cite{eikonal}. In other words, we witness
in our tree-level calculation the same mechanism at work in the loop
calculation of~\cite{ACV,us}: the exponentiation of the amplitude to
preserve the unitarity of the theory.

We are not interested here in this
part of the amplitude, which is simply the iteration of the four-point
amplitude. Instead, we want to study the truly non-linear effects.

\subsection{The Deflection Angle, I: Neutral String States}

\hspace*{.7cm} The $O (G_N^2) $ contribution to the eikonal phase
is the sum of two parts. The one from the Y-diagram (fig.3a)
given in eq.~(\ref{6a'}) and the second term in
eq.~(\ref{6b}) that is  somewhat hidden  as a sub-leading piece of
 the ladder diagram (figs. 3b and 3c).
This latter term was erroneously neglected in refs.~\cite{wrong,FI}.

Collecting these two terms together we have that
\beq
T_6  = \epsilon_{in} \cdot \epsilon_{out} \:
\frac{16 \kappa^4 M^{4} }
{{\bf q}_{\bot}^{2}\, {\bf q}_{1}^{2}\, {\bf q}_{2}^{2}}
\left[ E^2 {\bf q}_{\bot}^{2}
- ({\bf q}_{1} \cdot {\bf p}) ({\bf q}_{2} \cdot {\bf p}) \label{A6}
\right] \, ,
\eeq
and therefore
\bea
A^{(2)}_{fi} =
 & = &
 \frac{\kappa^4 M^2_{\displaystyle \star}}{{\bf q}^2_\bot}
\int \frac{\di ^3 {\bf q}_1\di ^3 {\bf q}_2}
{(2\pi)^6} \frac{1}{{\bf q}_{1}^{2} \,
{\bf q}_{2}^{2}} \nnu \\
& & \times
\left[ E{\bf q}_{\bot}^{2} - \label{delta-2}
\frac{({\bf q}_{1} \cdot {\bf p}) ({\bf q}_{2} \cdot {\bf p})}{E}
\right] (2\pi)^3 \delta^{(3)} ({\bf q} - {\bf q}_1 - {\bf q}_2 ) \, .
\eea
By means of the integrals in appendix C, we readily obtain that
\beq
 \delta^{(m=2)} (b)  =7\pi^2 G_{N}^{2} M^2_{\displaystyle \star} E
\int \frac{\di ^2 {\bf q}_\bot}{(2\pi)^2}
 e^{-i{\bf q}_\bot
\cdot {\bf b}} \frac{1}{|{\bf q}_\bot |} =
 \frac{7\pi}{2}  \frac{ G_N^{2} M^2_{\displaystyle \star} E }{b} \, .
\eeq

The deflection angle is  found by using~(\ref{deflection-angle}) and it
is
equal to
\beq
 \Delta \vartheta = \frac{4G_NM_{\displaystyle \star}}{b}
+ \frac{7\pi}{2} \left(
\frac{G_N M_{\displaystyle \star}}{b} \right) ^{2}
+ O ( G_N^3 ) \, .
\label{d2}
\eeq
(\ref{d2}) describes the scattering of a test particle in the
gravitational field of a string star made of massive and neutral states up to
$O ( G_N^2)$.

\subsection{The Deflection Angle, II: Charged String States}

\hspace*{.7cm} The case of massless
charged string states can be treated along similar
lines. Six massless vertices (the ones in (\ref{graviton}))
 replace the two massless and four massive vertices we used in the previous
section. Compactification on tori gives a mass to these scalar states
by requiring that they have at least one non-vanishing component of
momentum in one of the compact directions. The
ten-dimensional
momenta are now parametrized as follows:
\bea
 p_1' &=& (-\widetilde{E}_1 , -{\bf q}_1, -\xi_1 M) \nnu \\
  p_1 &=& (M , {\bf 0},+\xi_1 M ) \nnu \\
 p_2' & = & (-\widetilde{E}_2 , -{\bf q}_2, -\xi_2M)  \nnu \\
p_2 & = & (M , {\bf 0},+\xi_2 M ) \, ,\label{momenta'}
 \eea
while $p_{in}$ and $p_{out}$ are the same as before.
The last entry in (\ref{momenta'})  gives the
 non-zero momentum in one of the compact directions ,  $\xi_i = \pm 1$
 being the sign and $M$ is the four-dimensional mass.

The amplitude for this process can be computed and it is
identical to~(\ref{6}) except for the power of  minus six in the first
line
 being
replaced by a power of minus two. This is due to the different
mass-shell condition (that is, $\alpha 'p^2 = 0$ instead
of $-4$) now valid for the external
states. By proceeding in the computation
we obtain once again the expressions (\ref{6a}), (\ref{6b1}) and (\ref{6b2})
for the various pinching limits.

After subtracting the contribution
from the iteration of the ladder diagram which
gives rise again to
(\ref{iter}), we
can insert the values
of the momenta; these  are now given by (\ref{momenta'}), this being the
real difference with respect to section 2.3 above.
At the end we have
\beq
T_6  = \epsilon_{in} \cdot \epsilon_{out} \, \:
\frac{32 \kappa^4 M^{4} }
{{\bf q}_{\bot}^{2}\,{\bf q}_{1}^{2}\, {\bf q}_{2}^{2}} \left( 1 - \xi_1
\xi_2 \right) \left[ \frac{E^2}{2}{\bf q}_{\bot}^{2} -
({\bf q}_{1} \cdot {\bf p}) ({\bf q}_{2} \cdot {\bf p})
\right]  \label{aG}
\eeq
(recall that $\left( 1 - \xi_1
\xi_2 \right) ^2 = 2 \left( 1 - \xi_1
\xi_2 \right) $).

If the string star contains both positive and negative charges in such a
way that its total charge is zero, the term in (\ref{aG}) proportional to
$\xi_1\xi_2$ averages to zero. In this case the deflection angle can be
computed from
(\ref{T3}), (\ref{T}) and (\ref{deflection-angle}) and is given by
\beq
\Delta \vartheta = \frac{4G_NM_{\displaystyle \star}}{b} + 3\pi
\left( \frac{G_N M_{\displaystyle \star}}{b} \right) ^2
 + O ( G_N^3  ) \, .\label{def}
\eeq

If all the constituents carry charges of the same sign
(i.e., $\xi_1\xi_2 =1$), the string
star has a total charge $Q_{\displaystyle \star} =
\pm \sqrt{2}\kappa M_{\displaystyle \star}$
 and the second order
correction to the deflection angle vanishes.

\section{Field Theory Analysis}
\setcounter{equation}{0}

\hspace*{.7cm} In the calculation of the previous section, the
contribution of the various fields taking part in the interaction are
entwined together in the string amplitude. It is therefore useful
to reproduce our results by means of a
field theory analysis, in which each exchange of field contributes to an
independent Feynman diagram, in order to identify exactly which massless
fields are excited by the string matter. In section 4 this insight will
be used to obtain exact solutions from the low-energy effective action.

\subsection{Neutral String States}
\hspace*{.7cm} The massive scalars are neutral and therefore the
exchanged fields are only the graviton and the dilaton. This is
understood best from the ten-dimensional superstring from which one can
compute the amplitude for two massive string states to emit one (slightly
off shell) massless state.

Of the many massless states present in the superstring, only those in the
$NS \times \overline{NS}$ sector (that is, in the form (\ref{graviton})) have a
non-zero coupling to the massive state (\ref{scalar}), given by
\beq
2\kappa \epsilon_\mu \bar{\epsilon}_\nu p^\mu p^\nu \qquad
+ \qquad \left( {\mbox{longitudinal}
 \atop \mbox{piece}} \right) \, . \label{coupling}
\eeq
{}From (\ref{coupling}) we see that the coupling to a source with purely
four-dimensional momentum $p^\mu$ involves only the graviton and the
dilaton in four dimensions. In particular, the anti-symmetric tensor and
all states with polarizations pointing in compact directions decouple.

If we parametrize the gravitational field as
\beq
\sqrt{-g} g^{\mu\nu} \equiv \eta^{\mu\nu} + 2\kappa h^{\mu\nu} \, ,
\label{rule}
\eeq
there exist only two Feynman diagrams (fig.4). The relevant Feynman rules
are given in the appendix B.
\begin{figure}
\vspace{6cm}
\caption{Feynman diagrams for the neutral string states.}
\end{figure}

 The computation of the
three-graviton vertex is rather tedious; we write here  the final
result:
\beq
 T_{graviton}  = \epsilon_{in} \cdot \epsilon_{out} \, \:
\frac{16 \kappa^4 M^{4} }
{{\bf q}_{\bot}^{2}\, {\bf q}_{1}^{2}\, {\bf q}_{2}^{2}}
\: \left[E^2 {\bf q}_{\bot}^{2} - \frac{1}{2}
({\bf q}_{1} \cdot {\bf p}) ({\bf q}_{2} \cdot {\bf p})
\right] \, .\label{3g}
\eeq

This would be the only contribution in general relativity where only
gravitons are exchanged (fig.4a). Notice that this diagram contains
a contact term, the term without the pole
in ${\bf q}_\bot^2$, which in the string computation was hidden
in the ladder diagrams of fig.3b and 3c.

The corresponding diagram (fig. 4b) in which a dilaton is exchanged between the
two
massive states gives
\beq
 T_{dilaton}  =  -\epsilon_{in} \cdot \epsilon_{out} \, \:
\frac{8 \kappa^4 M^{4} }
{{\bf q}_{\bot}^{2}\, {\bf q}_{1}^{2}\, {\bf q}_{2}^{2}}
\:
({\bf q}_{1} \cdot {\bf p}) ({\bf q}_{2} \cdot {\bf p})
 \, ,
\eeq
and represents the coupling of gravity to the energy-momentum tensor of
the dilaton field.

We see that the string amplitude given by
 (\ref{A6}) originates
from these two diagrams. In particular, the dilaton gives the factor two
in front of the
\mbox{$({\bf q}_{1} \cdot {\bf p}) ({\bf q}_{2} \cdot {\bf p})$}
piece and does not contribute to the part proportional to $E^2$. The two
have opposite sign and therefore the dilaton contribution tends to make
the deflection angle smaller.

\subsection{Charged String States}
\hspace*{.7cm} The theory is more complicated in the case of  charged
string states acquiring a mass via toroidal compactification.
Let us assume for simplicity that the mass is generated by a non-zero value of
the momentum in only one (for instance, the fifth) direction.
As it can be readily understood from
(\ref{coupling}) such states couple to the {\em gravi}photon
(whose polarization is
$\epsilon_{\mu 5}$) and the {\em gravi}scalar ($\epsilon_{55}$)
as well as to the four-dimensional graviton. The Feynman rules and the precise
  field conventions are given in
 appendix B. Note that there is  no coupling to the
ten-dimensional dilaton
which couples only via a term proportional to the
ten-dimensional mass. Hence, we have
to consider the diagrams in fig.5  in addition
 to the contribution
of the graviton, eq.~(\ref{3g}).
\begin{figure}
\vspace{6cm}
\caption{Feynman diagrams for the charged string states.}
\end{figure}

The spin-1 exchange (fig.5a) gives
\beq
 T_{graviphoton}  =  - \epsilon_{in} \cdot \epsilon_{out} \, \:
\frac{16 \kappa^2 M^{2} Q_1Q_2 }
{{\bf q}_{\bot}^{2}\, {\bf q}_{1}^{2}\, {\bf q}_{2}^{2}}
\: \left[ \frac{E^2}{2} {\bf q}_{\bot}^{2} -
({\bf q}_{1} \cdot {\bf p}) ({\bf q}_{2} \cdot {\bf p})
\right] \, ,\label{1}
\eeq
where $Q_i = \sqrt{2}
\kappa \xi_i M$ are the charges. The spin-0 exchange (fig.5b) gives
\beq
 T_{graviscalar}  =  - \epsilon_{in} \cdot \epsilon_{out} \, \:
\frac{24 \kappa^4 M^{4} }
{{\bf q}_{\bot}^{2}\, {\bf q}_{1}^{2}\, {\bf q}_{2}^{2}}
\:
({\bf q}_{1} \cdot {\bf p}) ({\bf q}_{2} \cdot {\bf p})
 \, .\label{0}
\eeq
Together they yield the amplitude
\beq
 T_6  =  \epsilon_{in} \cdot \epsilon_{out} \, \:
\frac{32 \kappa^4 M^{4} }
{{\bf q}_{\bot}^{2}\, {\bf q}_{1}^{2}\, {\bf q}_{2}^{2}}
\left( 1 - \xi_1 \xi_2 \right)
\: \left[ \frac{E^2}{2}{\bf q}_{\bot}^{2}
- ({\bf q}_{1} \cdot {\bf p}) ({\bf q}_{2} \cdot {\bf p})
 \right] \, ,\label{aG'}
\eeq
 which agrees with (\ref{aG}).

\subsection{The Deflection Angle, III: General Relativity}

\hspace*{.7cm} For reference, we report here the  computation  of the
deflection angle in
field theory for general relativity. For a different but equivalent
computation, see~\cite{Duff}.
The amplitude (\ref{3g}) can be inserted in (\ref{T3}) instead of the
string result (\ref{A6}) or (\ref{aG}).
A straightforward computation gives
\beq
\Delta \vartheta =   \frac{4G_NM_{{\displaystyle \star}}}{b}
+ \frac{15\pi}{4} \left( \frac{G_N M_{{\displaystyle \star}}}{b} \right) ^{2}
+ O ( G_N^3 ) \, ,
\eeq
which agrees with the expansion to this order
 of the exact result~\cite{Darwin}.

The first order in $G_N$ term agrees with the string result by
construction (recall the way (\ref{k-G}) was defined). The $ O ( G_N^2 )$
terms are different. As a matter of fact, both string results are smaller
than the one in general relativity: for a star made out of neutral
massive string excitations, the numerical factor in front of the second
order term in (\ref{d2}) is $14\pi/4$
instead of
$15\pi/4$; whereas for the charged string star,
in the case where all constituents carry the same charge, this first non-linear
correction vanishes
(see section 2.4).
As we shall see, this decrease of the deflection angle reflects a
behavior of the gravitational field of string matter that is
 less
singular  at short
distances than the one of the Schwarzschild solution.

\section{The Exact Solutions}
\setcounter{equation}{0}

\hspace*{.7cm} Up to this point we have dealt directly with the string
amplitudes and we have analyzed
our results in terms of the various  fields
existing in the string theory.
The advantage of this approach is in the string taking
automatically into account all the fields that must be included.
On the other hand, the perturbative nature of the computation makes it
unsuitable for a discussion of strongly non-linear effects, such as the
existence or non-existence of horizons.

In this section we leave the string amplitude and move to the
effective action of the low-energy field theory, which can be  solved
exactly.

\subsection{Massive String Excitations}

The relevant part of the theory is described by the following
action (see appendix B for details):
\beq
S = \frac{1}{2\kappa^2} \int \di ^4 x \sqrt{-g} \: {\cal R}
-\frac{1}{2} \int \di ^4 x \sqrt{-g} \: g^{\mu\nu} \partial_\mu \phi
\partial_\nu \phi \label{action} \, ,
\eeq
where $\phi$ is the four-dimensional dilaton field. The action
(\ref{action}) is
equivalent\footnote{The mapping into the Brans-Dicke field
variables is given by $G_N\phi_{BD}=\exp (-2\kappa \phi /\sqrt{2})$, and
$g^{BD}_{\mu\nu} = g_{\mu\nu} /G_N\phi_{BD}$.}  to the
Brans-Dicke action~\cite{BD} for the particular value $-1$ of their
parameter $\omega$.

The equations of motion are
\beq
\partial_\mu \left( \sqrt{-g} \: g^{\mu\nu} \partial_\nu \phi \right) =
0\label{phi-eq}
\eeq
for the dilaton field, and Einstein's equations
\beq
{\cal R}_{\mu}^{\nu} - \frac{1}{2} \delta_{\mu}^{\nu} {\cal R} = \kappa^2 {\cal
T}_{\mu}^{\nu}\label{g-eq}
\eeq
for the graviton, where
\beq
{\cal T}_{\mu}^{\nu} =  \partial_\mu \phi
\partial^\nu \phi - \frac{1}{2} \delta_{\mu}^{\nu} g^{\alpha\beta}
\partial_\alpha \phi \partial_\beta \phi
\eeq
 is covariantly conserved.

To find the spherically symmetric solution we proceed in a manner similar
to the way the Schwarzschild solution is worked out in, for example,
ref.~\cite{LL}. Accordingly, we parametrize the metric tensor in
spherical coordinates
\beq
\di s^2 =  -e^{\nu (r)} \di t^2 + e^{\lambda (r)} \di r^2
+  r^2 \left( \di \theta^2 + \sin ^2 \theta \di \varphi^2 \right)
\label{metric}
\eeq

For this parametrization of the metric,
  the 11 and 00 component of (\ref{g-eq}) gives us the
 equations of motion
\bea
-\frac{1}{r^2} + e^{-\lambda} \left(\frac{\nu '}{r} + \frac{1}{r^2} \right)
&  = & \frac{\kappa^2}{2}e^{-\lambda}
\left(  \phi' \right) ^2 \label{11} \\
-\frac{1}{r^2} -e^{-\lambda} \left( \frac{\lambda '}{r} - \frac{1}{r^2} \right)
& = & -\frac{\kappa^2}{2}e^{-\lambda}
\left( \phi' \right) ^2\label{00}
\eea
while the other components yield no further information.
{}From (\ref{phi-eq}) we obtain the equation of motion for the dilaton
field:
\beq
 \left( e^{ \frac{\displaystyle \nu - \lambda}
{\displaystyle 2}} r^2  \phi' \right) ' =
r^2  \phi'' + \left( 2r + r^2 \frac{\nu ' - \lambda '}{r}
\right)  \phi' =  0\, .
\label{phi'}
\eeq
Everywhere $'$ denotes differentiation with respect to
the coordinate $r$.

It is now useful to introduce the two functions
\beq
V(r) \equiv \frac{\nu + \lambda}{2} \:\:\: \mbox{and} \:\:\:\:
W(r) \equiv \frac{\nu - \lambda}{2}
\eeq
and combine~(\ref{00}), (\ref{11}) and (\ref{phi'}) to get
\bea
  \phi'' +  \left( \frac{2}{r} + W' \right) \phi' & =  &0
 \label{a} \\
 \left( re^W \right) ' &  = & e^V  \label{4.13}
 \label{b} \\
V' & = & \frac{\kappa^2 r}{2}\left(  \phi' \right) ^2 \, .\label{c}
\eea

Eq.~(\ref{c}) can be differentiated to give, by means of (\ref{a}),
\beq
V'' + \frac{3}{r}V' + 2W' V' =  0  \label{eq-1}\, .
\eeq
Eq.~(\ref{eq-1}) can be integrated to find $V'$ as a function of $W$:
\beq
V' = \frac{c_0}{r^3} e^{-2W}   \, .  \label{Vprime}
\eeq
 Introducing the function
\beq
y \equiv r\, e^W \label{y} \, ,
\eeq
and using (\ref{4.13}), we obtain
\beq
\frac{y''}{ y'} = \frac{c_0}{r} \frac{1}{y^2}\label{diff-eq}
\eeq
where $c_0$ is a constant to be determined.
 Eq.~(\ref{diff-eq})
is now an ordinary differential equation to be solved.
This is possible by noting that~(\ref{diff-eq}) is obtained from the
simpler one,
\beq
ry' = -\frac{c_0}{y} + y + \Omega_0 \, ,\label{easy}
\eeq
by differentiation, $\Omega_0$ being another
integration constant. Therefore
\beq \left( \frac{r}{r_0} \right) =
\left( \frac{y-a}{y_0 - a} \right) ^{\frac{a}{a+b}}
\left( \frac{y+b}{y_0 + b} \right) ^{\frac{b}{a+b}}
  \, ,
\eeq
where
\bea
a & \equiv & \frac{1}{2} \left( \sqrt{\Omega_0 ^2 + 4c_0} - \Omega_0 \right)
\nnu \\
b & \equiv & \frac{1}{2} \left( \sqrt{\Omega_0 ^2 + 4c_0} + \Omega_0 \right)
\, .
\eea

Even though it seems that our solution
 depends on two new constants ($y_0$ and $r_0$)
the boundary condition on $W(r)$---that is,  the flatness of the metric
at infinity which implies  that $y \rightarrow r$ as $r \rightarrow
\infty$---tells us that
\beq
 (y_0 -a)^{\frac{a}{a+b}}
 (y_0 +b)^{\frac{b}{a+b}} = r_0  \, .
\eeq
Hence,
 we have
\beq
r = f(y) \equiv
\left( y - a \right) ^{\frac{a}{a+b}} \left( y+b \right) ^{\frac{b}{a+b}}
\label{fund-eq} \, ,
\eeq
which gives
$y$ implicitly as a function of $r$.

Finally, from (\ref{Vprime}), (\ref{diff-eq}) and  (\ref{fund-eq})
we have that
\bea
-g_{00} & = & \left( \frac{y-a}{y+b} \right) ^{\frac{b-a}{a+b}}
 \label{g00} \\
g_{11} & = & \frac{(y-a)(y+b)}{y^2} \nnu \, .
\eea

We discuss now this solution.

First of all, it is easy to see that the Schwarzschild solution is recovered
for
$c_0 = 0$ (that is,  $a=0$ and $b= \Omega_0$). In this case, (\ref{fund-eq})
reduces to
\beq
r = y + \Omega_0 \, .
\eeq
A distinctive feature of the Schwarzschild solution is  that negative
values of $y$ are possible. The horizon is where this
 change of sign of $y$ takes place. On the contrary,
the solution in our example of the string star has no horizon.
As it can be
 understood by studying graphically the solution
 of eq.~(\ref{fund-eq}) (fig.6),
 \begin{figure}
\vspace{6cm}
\caption{Relationship between the
radial coordinates $r$ and $y$.}
\end{figure}
 as $r$ goes to
zero, $y$ never crosses  over to  negative
values and reaches the limiting value of $a$ at $r=0$.
Accordingly,
$g_{00}  < 0 $ and $g_{11} > 0 $, as it is clear from~(\ref{g00}), and
therefore no component of the metric
ever changes sign.

In the opposite limit,
\beq
y \sim r - \Omega_0  \:\:\:\:\: \mbox{for}
\:\:\:  r \rightarrow \infty   \, .
\eeq
This behavior at large $r$ can be used to determine the constant
$\Omega_0$ by requiring that the metric goes into the one for Newton
theory:
\beq
-g_{00} \sim 1 - \frac{\Omega_0}{r} \equiv 1 -
\frac{2G_NM_{\displaystyle \star}}{r} \, ,
\eeq
from which
\beq
\Omega_0 = \frac{\kappa^2 M_{\displaystyle \star}}{4\pi} =
 2G_NM_{\displaystyle \star} \, .
\eeq

By inserting the solution~(\ref{g00}) for the metric into the equation of
motion
for the dilaton field (\ref{c}),
we find that
\beq
\phi = \sqrt{\frac{2c_0}{\kappa^2}}
\int^{r}_{\infty} \frac{\di \varrho}{\varrho y(\varrho)}
= \sqrt{\frac{2c_0}{\kappa^2}} \log \left(
\frac{y-a}{y+b} \right) ^{\frac{1}{a+b}} \label{fi} \, .
\eeq
The solution (\ref{fi}) can be used to determine the other unknown constant,
 $c_0$. In fact, at large $r$
\beq
\phi \sim -\sqrt{\frac{2c_0}{\kappa^2}}\frac{1}{r}\, ,
\eeq
which should correspond to the dilaton field of a collection
of $N$ massive strings at rest (see Feynman rules in appendix B)
\beq
-\frac{\kappa M_{\displaystyle \star}}{4\sqrt{2}\pi} \frac{1}{r} \, ,
\eeq
with the result that
\beq
c_0 = \left( \frac{\kappa^2 M_{\displaystyle \star}}
{8\pi} \right) ^2  = (G_NM_{\displaystyle \star})^2 \, . \label{4.29}
\eeq
Therefore, $a=G_NM_{\displaystyle \star}\left( \sqrt{2}-1\right)$ and
$b=G_NM_{\displaystyle \star}\left( \sqrt{2}+1\right)$.

A closed form of the solution\footnote{ The history of this solution is,
as far as we know it,
the following. It was written down  for the first time
 in~\cite{BD} and attributed to a
suggestion of C. Misner. In this reference the coordinates are the
so-called isotropic ones, which are related to the ours by
$\rho = y/2 + \sqrt{(y-a)(y+b)}/2 + (b-a)/4$.
 It was  given in~\cite{sol2} (without
reference to the previous work) in coordinates equivalent to ours. It
has been re-discovered independently in the present context
 by one of us (R.I.).}
can be found by a change of coordinates
in which $r$ is defined by (\ref{fund-eq}). In these coordinates
\beq
\di s^2 = -\left( \frac{y-a}{y+b } \right) ^{\frac{b-a}{a+b}}
\di t^2 + \frac{f^2(y)}{(y-a)(y+b)} \di y^2 + f^2(y) \left(
\di \theta^2 + \sin ^2 \theta \di \varphi ^2 \right) \, . \label{e-sol}
\eeq
The metric (\ref{e-sol}) belongs to the family of solutions related to
 the Brans-Dicke theory~\cite{BD}.
In our case, the parameters are fixed by
 matching the superstring perturbative theory.

The value $r=0$ (that is, $y=a$)  is the position of a
singularity of the curvature. In fact,
\beq
{\cal R} = 3 \frac{G_N^2 M_{\displaystyle \star}^2}
{r^2 (y-a)(y+b)} \, .
\eeq

Finally, as a consistency check we compare the exact solution to
 our perturbative computation of section 3  by expanding
the
solution (\ref{g00})  at large distances
\bea
-g_{00} & = & 1 - 2\frac{G_N M_{\displaystyle \star}}{r} +  O (G_N^3) \\
g_{11} & = & 1 + 2\frac{G_N M_{\displaystyle \star}}{r} + 3
\left( \frac{G_N M_{\displaystyle \star}}{r} \right) \label{gg}
^2 +  O (G_N^3) \, ,
\eea
and computing the deflection angle for a massless particle by
means of the formula~\cite{book}
\beq
\Delta \vartheta  =  2\, |\vartheta (r) - \vartheta _{\infty} | - \pi
\, ,
\eeq
where
\beq
\vartheta (r) - \vartheta _{\infty}=
\int_{r}^{\infty} g_{11}^{1/2} (r)
\left[ \left( \frac{r}{r_0} \right) ^2 \left( \frac{g_{00}(r_0)}{g_{00}(r)}
\right)
-1 \right] ^{-1/2} \frac{\di r}{r} \label{Angle} \, ,
\eeq
in which $r_0$ is the point of closest approach.

We thus find
\bea
\Delta \vartheta & = &
 4 \frac{G_N M_{\displaystyle \star}}{r_0} + \left( \frac{7\pi}{2} -4 \right)
\left( \frac{G_N M_{\displaystyle \star}}{r_0} \right) ^2
+ O (G_N^3)  \nnu \\
& & =
4 \frac{G_N M_{\displaystyle \star}}{b} +  \frac{7\pi}{2}
\left( \frac{G_N M_{\displaystyle \star}}{b} \right) ^2  + O(G_N^3)
\eea
in agreement with (\ref{d2}).

\subsection{Anti-Gravity}

\hspace*{.7cm} Next, we consider a string star made of states whose charge
and mass are given by a non-zero momentum in the compact fifth direction.
The exact solution of this theory gives rise to
 anti-gravity and it has been recently discussed
in~\cite{anti-gravity}.  It also corresponds to the extremal
value of
 $Q_{\displaystyle \star}/M_{\displaystyle \star}
= \pm \sqrt{2}\kappa$ (see section 2.4) discussed in~\cite{papers} for a
Maxwell field arising in the
compactification. We
report here only  the main results for ease of
 reference.
The relevant part of the effective
action involves only the graviton, one scalar
(the  gravi-scalar $\phi_{55} \equiv \delta$) and one vector
 (the  gravi-vector $A_\mu^5 \equiv A_\mu$)
 field  and is given in appendix B. Here we re-write the action
 introducing a scalar field $\sigma = -\sqrt{3} \log \delta /4$
  in such a way that it
 can be compared  directly to the similar action of ref.~\cite{papers}.
 Therefore, we have
\beq
S = \int \di ^4 x \sqrt{-g} \: \left\{ \frac{1}{2\kappa^2}
 {\cal R}
-\frac{1}{4} e^{-2\sqrt{3}\sigma} F_{\mu\nu} F^{\mu\nu}
-\frac{1}{\kappa^2} \left( \partial_\mu \sigma \right) ^2
\right\}\, , \label{Ag-action}
\eeq
 (\ref{Ag-action})
 gives rise to the following equations of motion:
\beq
{\cal R}_{\mu\nu} - \frac{1}{2} g_{\mu\nu} {\cal R} = \kappa^2 {\cal
T}_{\mu\nu}\label{g}
\eeq
for the graviton, with
\bea
{\cal T}_{\mu\nu} & = &  e^{-2\sqrt{3}\sigma} \left( F_{\mu\rho}F_{\nu}^{\
\rho}
-\frac{1}{4}g_{\mu\nu}F^{\rho\tau}F_{\rho\tau} \right)\nnu \\
& & +\frac{2}{\kappa^2} \left( \partial_\mu \sigma \partial_\nu
 \sigma
- \frac{1}{2} g_{\mu\nu} (\partial_\rho \sigma )^2
\right)
\eea
and
\beq
\nabla_\rho \left(e^{-2\sqrt{3}\sigma} F^{\rho\mu} \right)  \,
= 0
\eeq
for the gauge field. The scalar field obeys
\beq
 \nabla_{\rho}  \nabla^{\rho} \sigma +
\frac{\sqrt{3}}{4} \kappa^2 e^{-2\sqrt{3}\sigma}  F_{\mu\nu}F^{\mu\nu} = 0\, ,
\eeq
where $\nabla$ is the covariant derivative.

The simplest way of finding a solution to these equations
consists in going to the ten dimensional
 effective field theory where it is
 just a shock wave~\cite{anti-gravity}. In four space-time dimensions
 it reads as
\bea
\di s^{2} & = &
-\frac{\di t^2}{\left( \displaystyle 1+\frac{4G_N M_{\displaystyle \star}}{r}
 \right)^{1/2}}
 +
 \left( 1 + \frac{4G_N M_{\displaystyle \star}}{r} \right) ^{1/2}
 \left[\di r^2 +
 r^2 \left( \di \theta^2 + \sin^2 \theta \di \varphi^2 \right) \right] \nnu
\\
A_\mu & = & \frac{Q_{\displaystyle \star}}{4\pi r} \left(
1+\frac{4G_N M_{\displaystyle \star}}{r} \right) ^{-1}  \delta
^{0}_{\mu} \nnu \\
e^{-4 \sigma /\sqrt{3}} & = &1+\frac{4G_N M_{\displaystyle \star}}{r} \, ,
\label{4.42}
\eea
where the charge is $Q_{\displaystyle \star}=\pm \sqrt{2} \kappa
M_{\displaystyle \star}$.

The deflection angle for a massless and neutral test
particle can  be computed according to~(\ref{Angle}) and,
at large distances, we obtain
\beq
\Delta \vartheta = \frac{4G_N M_{\displaystyle \star}}{b} + O (G_N^3) \, ,
\eeq
in agreement with our perturbative computation in the case of a charged
string star.

(\ref{4.42}) coincides with the extremal solution of~\cite{papers} in
isotropic coordinates for the case in which their coupling $a$ of the scalar
field to the vector field is taken to be $\sqrt{3}$.

\section{Quantum Mechanics of the
Radial Motion of a Particle near the String Star}
\setcounter{equation}{0}

\hspace*{.7cm} In order to discuss the quantum mechanical problem of
the  motion of a particle in the field of
 the string star, we consider
 the  equation for the
particle's wave function $\psi$  in
the space-time~(\ref{gg}). For a massless particle (a small mass
would not alter the discussion), the wave equation follows from the action
\beq
S = \frac{1}{2}
\int \di t \: \di ^3 x \sqrt{-g} \left( |g^{00}| \partial_t \psi^*
\partial_t \psi - g^{ij} \partial_i\psi^*
\partial_j \psi \right) \, .
\eeq
The stationary and spherically symmetric solution
\beq
\psi = e^{iEt}\tilde{\psi} (r)
\eeq
is general enough because
a possible centrifugal barrier could only help
 in keeping the particle from falling into the origin.

The wave equation is obtained by varying $\tilde{\psi}$ in $S$. No boundary
terms
arise if
\beq
 \sqrt{-g} g^{11}\tilde{\psi}^* \partial_r\tilde{\psi} \rightarrow 0
 \label{bc}
\eeq
 at $r=\infty$ and at $r=0$.
 Otherwise, a boundary term would appear as a source (or a sink), and we
 would not have a source-free (or sink-free) wave equation.

 We thus have the equation
 \beq
 \frac{1}{\sqrt{-g}} \partial_r \left( \sqrt{-g} g^{11} \partial_r
 \tilde{\psi}
 \right) + |g^{00}| E^2 \tilde{\psi} = 0 \, , \label{s-eq}
 \eeq
 with the the boundary condition (\ref{bc}) at $r=0$.
We assume that at $\infty$ the condition
(\ref{bc}) is always satisfied  either because of
the exponential decay of $\tilde{\psi}$ for a bound state, or because we can
 form suitable wave-packets in the continuum spectrum.

For the metric~(\ref{gg}) we have, in the limit $r \rightarrow 0$
\bea
-g_{00} & \sim & \left( \frac{r}{a+b} \right)^{-1 +b/a} \\
g_{11} & \sim &  \left( \frac{a+b}{a} \right)^2
 \left( \frac{r}{a+b} \right)^{1 +b/a}\\
\sqrt{-g} & \sim & r^2 \sin \theta  \frac{a+b}{a}
 \left( \frac{r}{a+b} \right)^{b/a} \, .
\eea

The wave equation (\ref{s-eq}) is  of the Bessel type
\beq
\partial_r^2 \tilde{\psi} + \frac{1}{r} \partial_r \tilde{\psi} +
\frac{r^2}{a^2} E^2 \tilde{\psi} = 0 \, ,
\eeq
which admits   a regular solution:
\beq
\tilde{\psi} (r) \sim J_0 \left( \frac{E}{2a} r^2 \right) \, ,\label{J}
\eeq
where $J_0$ is the zero-order Bessel function which satisfies the
boundary condition (the other solution $Y_0 \left( \frac{E}{2a} r^2
\right)$ would not, and it must be discarded).

The same result holds also in the case of charged scalar  states,
where the regular solution is  proportional to
\beq
\frac{1}{\sqrt{r}} \: J_1 \left(  4E\sqrt{G_NM_{\displaystyle \star}r}
\right)
\eeq
 instead of (\ref{J}).

This way, the situation is  essentially similar to the  problem
 of the hydrogen atom in
quantum mechanics, where the
electron does not fall into the center because the indeterminacy principle
prevents it.

Therefore, the quantum mechanical behavior of a particle in the
space-time metric
of the string star is very mild, contrary to the case of the Schwarzschild
solution
where the particle falls into the horizon. In fact, in the case of the
Schwarzschild metric, a similar analysis leads to the conclusion that
\beq
\tilde{\psi} \simeq e^{\pm iEr_g \log (r-r_g)}
\eeq
where $r_g=2G_NM$ is the Schwarzschild radius for the star.
Clearly, the boundary condition at $r=r_g$ can never be satisfied. In terms of
the coordinate $\xi =r_g\log (r-r_g)$, the solution
for a particle starting out at $r=\infty$ represents a
 plane wave traveling towards and then through
 the horizon.

\section{Speculations}
\setcounter{equation}{0}

\hspace*{.7cm} In this final section
we would like to go back and
discuss some implications of the solution found in section 4.1 for the
gravitational field around a  string star
made from neutral string excitations. As we have seen,
the presence of the
 dilaton field changes drastically
the nature of the solution. The strength of the coupling of string matter to
 the dilaton field $\phi$ is
 fixed by string perturbation theory and is given by
the parameter $c_0$ of eq.~(\ref{4.29}) and thus the parameter $a$ of
eq.~(\ref{easy}). For $c_0=a=0$ we recover the standard Schwarzschild
solution.

It is interesting  that even considering $c_0$ as a free parameter and
taking it to be very small, the solution remains of the same kind,
namely free of horizons and deviating
 from the Schwarzschild one  near the would-be horizon.
Therefore, in this sense, the
Schwarzschild metric appears to be unstable with respect to the presence
of the dilaton field.

In order to understand better this point, it is useful to consider the equation
for a static scalar field $\phi$ in the background of the Schwarzschild
metric. For $x\equiv r - r_g \rightarrow 0$, that is near the horizon at
$r_g$,
the equation becomes
\beq
\partial^2 \phi + \frac{1}{x} \partial \phi -\frac{m^2r_g}{x} \phi = 0
\label{*}
\eeq
whose solution is $\phi \approx \log x$ for $x \rightarrow 0$;
 this is the only
solution if we require that $\phi$ does not grow exponentially for $x
\rightarrow \infty$.

Therefore the energy-momentum tensor of $\phi$  diverges for  $x
\rightarrow 0$ producing a singularity in $\cal R$. Hence,
 by computing the
back-reaction of the field $\phi$
 on the metric through the
Einstein equations, one expects to get a real singularity without horizon.
This is indeed what occurs in our solution of
section 4.1.

As we have already mentioned, similar solutions of the low-energy action
have been recently discussed in the literature~\cite{papers}, where there is a
dilaton field and still the horizon is present. In these cases however
the electric or magnetic charge is also different from zero  and the
coupled equations for the dilaton and the Maxwell field prevent
singularities at the horizon. This is of course not what we find for the
case of the neutral string excitation. Similarly,  the anti-gravity case
we considered corresponds to an extremal solution where there is a Maxwell
field but
still there is no horizon.

It is also interesting to notice that a mass term for the dilaton field
does not alter qualitatively the above picture; we see from eq.~(\ref{*}) that
for
 $x
\rightarrow 0$ the behavior is the same as for $m=0$, consistent with
the fact that near the horizon any  field
independent of the external time coordinate $t$ is
propagating almost at the speed of light.

Accordingly, even though we discussed
 a scenario where compactification has occurred
without the dilaton field acquiring a mass, one can speculate that the
solution of section 4.1 might be of relevance even
 in the more realistic case in which the dilaton field becomes
massive.
For $r-r_g \gg 1/m_{dilaton}$ the solution will approach the
Schwarzschild one, but for $r-r_g \ll 1/m_{dilaton}$ we expect the
solution to be indistinguishable from the one discussed in section 4.1.
At  distances of the order of the
Planck length  however interactions of higher orders in
$\alpha '$~\cite{stringhe} become important and a classical solution
should merge into the full quantum gravity picture, a task beyond the
scope of the present work.

\appendix
\section{Four-Point Amplitude}
\setcounter{equation}{0}

\hspace*{.7cm} In this appendix we show how the six-point
string amplitude we have
derived within the framework of the covariant loop calculus, can also be
 obtained by starting out from the four-point graviton-graviton
amplitude. This is given in an explicit way in ref.~\cite{ST}
as
\bea
T_4^{tree} & = & - \frac{(\alpha')^3}{4} \kappa^2 \frac{\Gamma(-\alpha' s/4)
\Gamma(-\alpha' t/4) \Gamma (-\alpha' u/4)}{\Gamma(1+\alpha' s/4)
\Gamma(1+\alpha' t/4) \Gamma(1+\alpha' u/4)} \nnu \\
& & \times \epsilon^{(1)}_{i_1 j_1} \epsilon^{(2)}_{i_2 j_2}
\epsilon^{(3)}_{i_3 j_3} \epsilon^{(4)}_{i_4 j_4}
{\cal K}^{i_1 i_2 i_3 i_4 }{\cal K}^{j_1 j_2 j_3 j_4 } \nnu \\
& = &
\frac{16\kappa^2}{stu} \epsilon^{(1)}_{i_1 j_1}
\epsilon^{(2)}_{i_2 j_2}\epsilon^{(3)}_{i_3 j_3}\epsilon^{(4)}_{i_4 j_4}
{\cal K}^{i_1 i_2 i_3 i_4 }{\cal K}^{j_1 j_2 j_3 j_4 }
+ O(\alpha')   \label{A1}
\eea
where $s,t,u$ are the usual Mandelstam variables. If we take legs two and
three to represent the incoming and outgoing gravitons and legs one and four
to represent the virtual gravitons exchanged with the string star, as shown
in fig.(7),
 we have
\bea
s & = &-(k_1 + k_2 )^2 \simeq 2 {\bf p} \cdot {\bf q}_1 \nnu \\
t & = & -(k_2 + k_3 )^2 \simeq -{\bf q}^2_{\bot} \nnu \\
u & = & -(k_1 + k_3 )^2 \simeq -s \, .
\eea
In order to close the two off-shell legs $(n=1,4)$ and form an on-shell
six-point amplitude we make the substitution
\beq
\epsilon^{(n)}_{i_n j_n} \rightarrow
2 \kappa \frac{\Delta_{\mu}^{(n)} \Delta_{\nu}^{(n)}}{{\bf q}_n^2}
\ \ \ \ \ \mbox{for} \ \ n=1,4    \label{A3}
\eeq
\begin{figure}
\vspace{6cm}
\caption{Four graviton vertex.}
\end{figure}
where
\beq
\Delta_{\mu}^{(n)} = (M, {\bf q}_n/2) \simeq (M, {\bf 0})
\eeq
is the average momentum of the string matter inside the star. The normalization
of~(\ref{A3}) is dictated by factorization.

 ${\cal
K}$ is given by a rather lengthy expression that contains the kinematical part
of
the graviton-graviton amplitude, it can be found
 in full in eq.~(2.27) of
ref.~\cite{ST}. Of this, we only need the piece proportional to
\beq
\delta ^{i_2 i_3} \delta ^{j_2 j_3}
\eeq
which gives rise to the part of the amplitude proportional to
 $\epsilon_{in} \cdot \epsilon_{out}$.

This leaves us with
\bea
{\cal K}^{i_1 i_2 i_3 i_4 }{\cal K}^{j_1 j_2 j_3 j_4 }
& \simeq & \delta ^{i_2 i_3} \delta ^{j_2 j_3} \left\{
-\frac{1}{4} us \delta^{i_1 i_4} + \frac{s}{2} k_3^{i_1} k_2^{i_4}
+ \frac{u}{2} k_2^{i_1} k_3^{i_4} \right\}  \nnu \\
& & \times \left\{
-\frac{1}{4} us \delta^{j_1 j_4} + \frac{s}{2} k_3^{j_1} k_2^{j_4}
+ \frac{u}{2} k_2^{j_1} k_3^{j_4} \right\} \, .
\eea
Inserting this into (\ref{A1}) and replacing
$\epsilon^{(1)}, \epsilon^{(4)}$ according to~(\ref{A3})  we find
\beq
T_6 = \frac{4\kappa^4}{stu} \frac{\epsilon_{in} \cdot \epsilon_{out}}{
{\bf q}_1^2 {\bf q}_2^2} \left( -us \Delta^{(1)} \cdot \Delta^{(2)}
+ 2t \Delta^{(1)} \cdot p_{in} \Delta^{(2)} \cdot p_{in} \right)^2 \, .
\eeq
The square of the first term in the bracket yields
\beq
-\frac{4\kappa^4 s^2}{t} \frac{\epsilon_{in} \cdot
\epsilon_{out}}{{\bf q}_1^2{\bf
 q}_2^2} \left( \Delta^{(1)} \cdot \Delta^{(2)} \right) ^2
= \frac{16 \kappa^4 M^4 \epsilon_{in} \cdot
\epsilon_{out}}{ {\bf q}^2_{\bot}
{\bf q}_1^2{\bf
 q}_2^2} ({\bf p} \cdot {\bf q}_1 )^2\, ,
\eeq
whereas the double product gives the contact term
\beq
-\frac{16\kappa^4 \epsilon_{in} \cdot
\epsilon_{out}}{{\bf q}_1^2{\bf
 q}_2^2} \left( \Delta^{(1)} \cdot \Delta^{(2)} p_{in} \cdot  \Delta^{(1)}
p_{in} \cdot  \Delta^{(2)} \right)  = \epsilon_{in} \cdot
\epsilon_{out} \frac{16\kappa^4 M^4 E^2}{
{\bf q}_1^2 {\bf  q}_2^2}\, ,
\eeq
in agreement with our previous results (\ref{A6}).

\section{Feynman rules}
\setcounter{equation}{0}

\hspace*{.7cm}In this appendix we present our definitions
for the various field variables which
couple to the string matter in $D=10$ and $d=4$ dimensions.

In $D=10$, only the graviton and the dilaton fields are generated
(see, eq.~(\ref{coupling})). The relevant part of the superstring effective
action
is therefore ~\cite{GSW}
\bea
\hat{S} & =  & \frac{1}{2\hat{\kappa}^2} \int \di ^D \hat{x} \sqrt{-\hat{g}}
\:e^{-2\hat{\Phi}}
\left( \hat{R} + 4 \dif_{\hat{\mu}} \hat{\Phi} \dif ^{\hat{\mu}} \hat{\Phi}
\right) \nnu \\
& & -\frac{1}{2} \int \di ^D \hat{x} \sqrt{-\hat{g}}
\:e^{-2\hat{\Phi}}
\left( \dif_{\hat{\mu}} \hat{B} \dif^{\hat{\mu}} \hat{B}
+ \hat{M}^2 \hat{B}^2  \right) \, , \label{B1}
\eea
where $\hat{\Phi}$ is the $10$-dimensional dilaton field and we adopt the
convention that all $10$-dimensional objects carry a hat. In
eq.(\ref{B1}) we represented the string matter by a field $\hat{B}$.
That this reproduces the correct coupling of string matter to graviton and
dilaton is seen by performing the rescalings
\bea
\hat{g}_{\hat{\mu} \hat{\nu}} & \rightarrow & e^{4\hat{\Phi}/(D-2)}
\hat{g}_{\hat{\mu} \hat{\nu}}  \nnu \\
\hat{\Phi} & \rightarrow & \sqrt{D-2}\ \frac{\hat{\kappa}}{2} \ \hat{\Phi}
\label{B2}
\eea
in terms of which the effective action (\ref{B1}) assumes the canonical form
\beq
\hat{S}  =  \int \di^D \hat{x} \sqrt{-\hat{g}} \left\{
\frac{1}{2\hat{\kappa}^2} \hat{R} - \frac{1}{2} \dif_{\hat{\mu}} \hat{\Phi}
\dif^{\hat{\mu}} \hat{\Phi} - \frac{1}{2} \dif_{\hat{\mu}} \hat{B}
\dif^{\hat{\mu}} \hat{B}
-\frac{1}{2}\hat{M}^2
e^{2\hat{\kappa} \hat{\Phi}/\sqrt{D-2}} \hat{B}^2 \right\}
\label{B3}
\eeq
As we perform a toroidal compactification down to $d=4$, the internal
components
of the metric give rise to graviphoton fields $A_{\mu}^{\alpha}$ and
graviscalar fields $\phi_{\alpha \beta}$, where $\mu,\nu = 1, \ldots ,d$
denote $d$-dimensional space-time indices and $\alpha,\beta = 1, \ldots,
D-d$ denote the internal ones. We parametrize the $10$-dimensional metric
as follows ~\cite{Scherk/Schwarz,anti-gravity}
\beq
\hat{g}_{\hat{\mu} \hat{\nu}} = \left( \begin{array}{cc}
\delta^{\gamma} g_{\mu \nu} + 2 \kappa^2 A_{\mu}^{\alpha} A_{\nu}^{\beta}
\phi_{\alpha \beta} & -\sqrt{2} \kappa A_{\mu}^{\beta} \phi_{\alpha \beta}
\\ -\sqrt{2} \kappa A_{\nu}^{\alpha} \phi_{\alpha \beta} &
\phi_{\alpha \beta}
\end{array}
\right)
\label{B4}
\eeq
where $\delta \equiv \det \phi_{\alpha \beta}$ and $\gamma \equiv -1/(d-2)$.
In terms of these fields the effective action (\ref{B3}) becomes
\bea
S & = & \int \di^d x \sqrt{-g} \left\{ \frac{1}{2\kappa^2}R
-\frac{1}{2} \dif_{\rho} \Phi \dif^{\rho} \Phi
-\frac{1}{4} \delta^{-\gamma} F^{\mu \nu \alpha} F^{\beta}_{\mu \nu}
\phi_{\alpha \beta} \right. \nnu \\
 & & -\frac{1}{8\kappa^2} \phi^{\alpha_1 \alpha_2} \phi^{\beta_1 \beta_2}
\dif_{\rho} \phi_{\alpha_1 \beta_1} \dif^{\rho} \phi_{\alpha_2 \beta_2}
- \frac{1}{8 \kappa^2 (d-2)} \dif_{\rho} \log \delta \ \dif^{\rho} \log \delta
\nnu \\
& & -\frac{1}{2} g^{\mu \nu} ( \dif_{\mu} + i \sqrt{2} \kappa p_{\alpha}
A_{\mu}^{\alpha} ) B \ (\dif_{\nu} - i \sqrt{2} \kappa p_{\beta}
A_{\nu}^{\beta}) B \nnu \\
& & \left.
-\frac{1}{2} \phi^{\alpha \beta} p_{\alpha} p_{\beta} B^2 \delta^{\gamma}
-\frac{1}{2} \hat{M}^2 \delta^{\gamma} e^{2\kappa \Phi/\sqrt{D-2}} B^2 \right\}
\label{B5}
\eea
where we have introduced compact momenta $p_{\alpha}$ for the field $B$. We
have also introduced the $d$-dimensional gravitational coupling, $\kappa$,
and $d$-dimensional canonical fields, $B$ and $\Phi$, by rescaling the
$D$-dimensional ones by the appropriate power of the coordinate volume
of the compact space.

We can now distinguish between two cases:

{\bf 1} . If the string matter is massless in $D=10$ (i.e. $\hat{M}=0$) the
$10$-dimensional dilaton decouples.

Assuming for simplicity that only one of the compact momenta is non-zero,
$p_{\beta} \equiv M \delta_{\beta}^{\alpha}$, we see from (\ref{B5}) that
only one graviphoton field, $A_{\mu}^{\alpha} \equiv A_{\mu}$, and one
graviscalar field, $\phi_{\alpha \alpha} \equiv \delta$, is excited.

The effective action is reduced to (taking $d=4$):
\bea
S & = & \int \di^4 x \sqrt{-g} \left\{ \frac{1}{2\kappa^2} R
- \frac{1}{4} \delta^{3/2} F_{\mu \nu} F^{\mu \nu} \right. \nnu \\
& & \left. -\frac{3}{16 \kappa^2} \dif_{\rho} \log \delta \dif^{\rho}
\log \delta - \frac{1}{2} \dif_{\rho} B \dif^{\rho} B
-\frac{1}{2} m^2 \delta^{-3/2} B^2 \right\} \label{B6}
\eea

{\bf 2}. If the string matter is massive in $D=10$ ($\hat{M} \neq 0$) but does
not carry any compact momentum, the only relevant fields besides the
graviton are the $10$-dimensional dilaton $\Phi$ and the determinant
of the internal metric, $\delta$. The ``transverse'' graviscalar fields
$\phi^T_{\alpha \beta} \equiv \phi_{\alpha \beta}/ \delta^{1/(D-d)}$
decouple.

If we define the $4$-dimensional dilaton field by
\beq
\phi \equiv \sqrt{\frac{d-2}{D-2}} \Phi - \frac{1}{2\kappa} \frac{1}{
\sqrt{d-2}} \log \delta  \label{B7} \, ,
\eeq
the orthogonal combination decouples and the effective action is reduced
to
\bea
S & = & \int \di^d x \sqrt{-g} \left\{ \frac{1}{2\kappa^2} R
- \frac{1}{2} \dif_{\rho} \phi \dif^{\rho} \phi \right. \nnu \\
& & \left. -\frac{1}{2} \dif_{\rho} B \dif^{\rho} B
-\frac{1}{2} M^2 e^{2\kappa \phi/\sqrt{d-2}} B^2 \right\}
\label{B8}
\eea
Notice that the string matter field $B$ does not appear in the Einstein
equations (\ref{g-eq}) or (\ref{g}).
This is because this field represents the distribution of string matter
at the origin, whereas we solve the equations of motion for the massless
fields outside this matter distribution.
Like in the case of the usual Schwarzschild
solution, reference to the matter source is only made in determining
the asymptotic behavior at large distance for the massless fields.
Accordingly, we only need
the linear coupling of the graviton and the dilaton to the string matter
field $B$.

If we define the graviton and graviscalar perturbations by (\ref{rule}) and
\beq
\phi_{\alpha \beta}  \equiv  \delta_{\alpha \beta} + 2\kappa h_{\alpha
\beta}  \label{B9}
\eeq
one can derive from (\ref{B5}) the set of relevant Feynman rules for the
massless fields.
For the {\bf graviton}:
\vspace*{1cm}
\beq
\hspace*{5cm} \frac{1}{2q^2} (\eta^{\mu_1 \mu_2} \eta^{\nu_1 \nu_2}
+ \eta^{\mu_1 \nu_2} \eta^{\nu_1 \mu_2} - \eta^{\mu_1 \nu_1} \eta^{\mu_2
\nu_2})
\eeq
\vspace*{1cm}
\bea
\hspace*{5cm} & &
\frac{\kappa}{2} \mbox{sym} P_6 \left\{ -4 q_2 \cdot q_3 \eta_{\mu_2 \mu_3}
\eta_{\mu_1 \nu_2} \eta_{\nu_1 \nu_3}\right.  \\
& & + \frac{4}{d-2} q_2 \cdot q_3 \eta_{\mu_2 \nu_2} \eta_{\mu_3 \mu_1}
\eta_{\nu_3 \nu_1}
+ 2 q^2_{\mu_1} q^3_{\nu_1} \eta_{\mu_3 \mu_2}
\eta_{\nu_2 \nu_3} \nnu \\
& & \left. -\frac{2}{d-2} q^2_{\mu_1} q^3_{\nu_1} \eta_{\mu_2 \nu_2}
 \eta_{\mu_3 \nu_3}
+ 4 q^2_{\mu_3} q^3_{\nu_2} \eta_{\mu_2 \mu_1}
\eta_{\nu_3 \nu_1} \right\} \nnu
\eea
where ``sym $P_6$'' implies: 1) symmetrization in $\mu_1 \leftrightarrow
\nu_1$, $\mu_2 \leftrightarrow
\nu_2$ and $\mu_3 \leftrightarrow
\nu_3$ with unit weight. 2) Adding all permutations of legs 1, 2 and 3.

For the {\bf graviphoton}:
\vspace*{1cm}
\beq
\hspace*{5cm} \frac{\eta_{\mu \nu}}{q^2} \delta^{\alpha \beta}
\eeq
\vspace*{1cm}
\bea
\hspace*{5cm} & &
-\kappa \left\{ p_1 \cdot p_2 ( \eta^{\mu \nu} \eta_{\rho \sigma}
- \delta^{\nu}_{\sigma} \delta^{\mu}_{\rho} - \delta^{\nu}_{\rho}
\delta^{\mu}_{\sigma}) - (p^1_{\rho} p^2_{\sigma} + p^1_{\sigma} p^2_{\rho})
\eta^{\mu \nu} \right. \nnu \\
& & -p_1^{\nu} p_2^{\mu} \eta_{\rho \sigma}
+ p_1^{\nu} (p^2_{\sigma} \delta^{\mu}_{\rho} + p^2_{\rho}
\delta^{\mu}_{\sigma} ) \nnu \\
& &\left.+ p_2^{\mu} (p^1_{\rho} \delta^{\nu}_{\sigma}
+ p^1_{\sigma} \delta^{\nu}_{\rho} ) \right\}
\eea
\bigskip
For the {\bf graviscalar}:
\vspace*{1cm}
\beq
\hspace*{5cm} \frac{1}{2q^2} (\delta_{\alpha_1 \alpha_2} \delta_{\beta_1
\beta_2}
+ \delta_{\alpha_1 \beta_2} \delta_{\beta_1 \alpha_2} -
\frac{2}{D-2} \delta_{\alpha_1 \beta_1} \delta_{\alpha_2
\beta_2} )
\eeq
\vspace*{1cm}
\bea
\hspace*{5cm} & &
 \frac{\kappa}{2} (p^1_{\mu} p^2_{\nu} + p^1_{\nu} p^2_{\mu}) \cdot
(\delta^{\alpha_1 \alpha_2} \delta^{\beta_1 \beta_2}
+ \delta^{\alpha_1 \beta_2} \delta^{\beta_1 \alpha_2} \nnu \\
& & + \frac{2}{d-2} \delta^{\alpha_1 \beta_1} \delta^{\alpha_2 \beta_2} )
\eea
\bigskip

For the {\bf $d$-dimensional dilaton}:
\bigskip
\beq
\hspace*{5cm} \frac{1}{q^2}
\eeq
\bigskip

The coupling to string matter can also be read off from (\ref{B5}).
The graviton has a universal coupling to energy-momentum
\vspace*{1cm}
\beq
\hspace*{5cm}
 \kappa \left\{
p_{\mu}^1 p_{\nu}^2 + p_{\nu}^1 p_{\mu}^2 - \frac{2}{d-2}M^2
\eta_{\mu \nu} \right\}
\eeq
\vspace*{1cm}
where $M$ is the mass of the string state in $d$ dimensions.

The graviscalars and graviphotons couple to a string state with compact
momentum $p_{\alpha}$, mass $M^2 = \delta^{\alpha \beta} p_{\alpha} p_{\beta}$
and charge $q_{\alpha} = \sqrt{2} \kappa p_{\alpha}$ in the following way:
\vspace*{1cm}
\beq
\hspace*{5cm}
  q_{\alpha} (p_2^{\mu} - p_1^{\mu})
\eeq
\vspace*{1cm}
\beq
\hspace*{5cm}
\frac{1}{\kappa} q_{\alpha} q_{\beta} + \frac{2\kappa}{d-2} m^2
\delta_{\alpha \beta}
\eeq
\vspace*{1cm}
The coupling of the $d$-dimensional dilaton to a string state massive
already in $D$ dimensions is
\vspace*{1cm}
\beq
\hspace*{5cm}
 - \frac{2\kappa}{\sqrt{d-2}} M^2
\eeq
\vspace*{1cm}
The coupling of the graviton to the dilaton is given by
\vspace*{1cm}
\beq
\hspace*{5cm}
 \kappa \left\{ p^1_{\mu} p^2_{\nu} +
 p^1_{\nu} p^2_{\mu} \right\}
\eeq
\vspace*{1cm}

\section{Useful Integrals}
\setcounter{equation}{0}

\hspace*{.7cm} Integration of Koba-Nielsen variables:
\bea
 \int \di ^2 A | A | ^{-2 + \alpha ' p_1 \cdot p_2} &=&
\frac{2}{\alpha ' p_1 \cdot p_2} \int \di ^2 A \bar{\partial}_A \left(
\bar{A}^{\alpha ' p_1 \cdot p_2 /2} A^{-1+ \alpha ' p_1 \cdot p_2 /2}
\right) \nnu \\
& & \longrightarrow
 \frac{2\pi}{\alpha ' p_1 \cdot p_2 }\:\:\: \mbox{for}
 \:\:\:  \alpha ' \rightarrow 0 \, .
\eea

 Momenta integrals\footnote{These integrals,with the exception
of (C.3), are ultraviolet divergent and are
most conveniently done by
 dimensional regularization. An
 ultraviolet cutoff is physically provided by the star wave function, see
 section 2 and Appendix D.}:
 \bea
 \frac{1}{{\bf q}^2} \int \frac{ \di ^3 {\bf q}_1}{(2\pi)^3}
 \frac{{\bf q}_1 \cdot
 {\bf q}_2}{{\bf q}_1^2 \:{\bf
 q}_2^2} &  = & \frac{1}{16 \sqrt{{\bf q}^2}}\\
\frac{1}{{\bf q}^2} \int \frac{ \di ^3 {\bf q}_1}{(2\pi)^3}
\frac{ {\bf q}^2}{{\bf q}_1^2{\bf
 q}_2^2} & = & \frac{1}{8\sqrt{{\bf q}^2}}\\
 \frac{1}{{\bf q}^2} \int \frac{ \di ^3 {\bf q}_1}{(2\pi)^3}
\frac{{\bf q}_1 ^2+ {\bf q}_2 ^2 }{{\bf q}_1^2{\bf
 q}_2^2} & = & 0\\
 \frac{1}{{\bf q}^2} \int \frac{ \di ^3 {\bf q}_1}
{(2\pi)^3} \frac{{\bf p}_i \cdot {\bf q}_1 {\bf p}_j \cdot {\bf q}_2}{{\bf
q}_1^2{\bf
 q}_2^2} &
 =& \frac{1}{64} \left(  \frac{{\bf p}_i \cdot {\bf q}
  \:{\bf p}_j \cdot {\bf q}}{ ({\bf q}^2 )^{3/2}} +
   \frac{{\bf p}_i \cdot
  {\bf p}_j }{ \sqrt{{\bf q}^2}} \right)  \, .
 \eea

D-dimensional Fourier Transform:
\beq
\int \frac{\di ^D q}{(2\pi)^D} F(q) e^{-i q \cdot r} =
\frac{1}{(2\pi)^{D/2}} \frac{1}{r^{(D-2)/2}}
\int_{0}^{\infty} q^{D/2} F(q) J_{(D-2)/2}(qr) \di q \, ,
\eeq
where the last integral can be found by using~\cite{Tables}
\beq
\int_{0}^{\infty} x^\mu J_\nu (ax) \di x = 2^\mu a^{-\mu -1}
\frac{\Gamma \left( \frac{1}{2} + \frac{\mu + \nu}{2} \right)}
{\Gamma \left( \frac{1}{2} + \frac{\nu - \mu}{2} \right)} \, .
\eeq
For instance, we have
\bea
\int \frac{\di ^2 {\bf q}_{\bot}}{(2\pi)^2} e^{-i{\bf q}_{\bot} \cdot {\bf b}}
\frac{1}{{\bf q}_{\bot}^2} & = & -\frac{1}{2\pi} \log b \\
\int \frac{\di ^2{\bf q}_{\bot}}{(2\pi)^2} e^{-i{\bf q}_{\bot} \cdot {\bf b}}
\frac{1}{|{\bf q}_{\bot}|} & = & \frac{1}{2\pi b}\, .
\eea

\section{Kinematics}

\hspace*{.7cm}
The kinematics for the amplitude (\ref{A4}) in which
 only one graviton is exchanged (fig.1b)
 is simple.
 The massless particle comes in with a large energy $E$.
In the Regge regime we are interested in, the square of the
exchanged momentum $q^2$ is fixed and  such that
also $q^2/M^2 \ll 1$.

In the laboratory frame in which the $N$ massive particles are initially at
rest,
the external momenta (fig.2) are parametrized in terms of
 \bea
 p_{in} = (E, {\bf p}) & ; & p_{out} = (-E' , {\bf q} - {\bf p}) \nnu \\
 p_1 & = & (M, {\bf 0}) \nnu \\
 p_1' &=& (-\widetilde{E}, -{\bf q}) \, ,
 \eea
 where $E=|{\bf p}|$,
 \beq
 E' = \sqrt{({\bf q} - {\bf p})^2} \simeq E -\frac{ {\bf p} \cdot {\bf q}}
 {E} \equiv E - q_L
 \eeq
 and
 \beq
 \widetilde{E} = \sqrt{M^2 + {\bf q}^2} \simeq M + \frac{{\bf q}^2}{2M}\, .
 \eeq

 From the conservation of the energy,
\beq
\mbox{q}_0 = E - E' = \widetilde{E} - M \, ,\label{energy}
\eeq
  we obtain that
 \beq
 \mbox{q}_0 = \mbox{q}_L  = \frac{{\bf q}^2}{2M} \label{q0-ql} \,
 .\label{qua}
 \eeq
 Eq.~(\ref{qua}) can be solved to give
 \beq
 \mbox{q}_0 = \mbox{q}_{L} =
\frac{{\bf q}_{\bot}^2}{2M} \left( 1 + O\left(
 \frac{{\bf q}_{\bot}^2}{M^2} \right) \right) \, ,
 \eeq
where $\mbox{q}_{L}$ is the component of ${\bf q}$ in
the direction of the incoming graviton and  ${\bf q}_{\bot}$
is the transverse part.
Hence, we have
 zero energy transfer in the limit $|{\bf q}_{\bot}| / M
\rightarrow 0$.

 From (\ref{q0-ql}) also follows that
\beq
q^2 =  {\bf q}_{\bot}^{2}  \, ,
\eeq
so that the square of the transferred momentum comes
from the transverse part only.

The amplitude (\ref{6}) in which two
massive string states partake in the interaction
gives a slightly more complicated kinematics (see
fig1.b and fig.2).
This time the external momenta are given by (\ref{momenta}).

Energy conservation yields
\beq
\mbox{q}_{0} = E - E' = q_L \, , \label{k2}
\eeq
and
\beq
q_0 =  (\widetilde{E}_1 -M ) +
(\widetilde{E}_2 - M) \simeq
\frac{{\bf q}^{2}_{1} + {\bf q}^{2}_2}
{2M}  + \:\: \cdots  \, .\label{k1}
\eeq
Eq.~(\ref{k1}) and (\ref{k2}) give that
\beq
\frac{\mbox{q}^{1}_{L} + \mbox{q}^{2}_{L}}{M} \simeq
\frac{{\bf q}^{2}_{1} + {\bf q}^{2}_2}
{2M^2} \ll 1\, .  \label{region2}
\eeq
Eq.~(\ref{region2}) together with the identity
\beq
{\bf q}^{1}_{\bot} +{\bf q}^{2}_{\bot} = {\bf q}_{\bot}
\eeq
give us two kinematical constraints in the region over which
 the final momenta of the massive
scalars are integrated.
In particular we are interested in quasi-elastic scattering where only a
relatively small excitation energy is imparted to the scatterer, and in which
therefore q$_0 \leq \Lambda $. This cutoff is taken automatically into
 account  by dimensional regularization which put the infinite part equal
to zero.

\newpage
\renewcommand{\baselinestretch}{1}

\end{document}